\documentclass{aa}  

\usepackage{graphicx}
\usepackage{txfonts}
\usepackage{lipsum}
\usepackage{subcaption}         
\usepackage{lscape}             
\usepackage{placeins}           

\usepackage{xcolor}
\usepackage{longtable}
\usepackage[normalem]{ulem}    
\usepackage{hyperref}

\usepackage{todonotes}
\newcommand{\incom}[1]{}

\begin{document}

   \title{Results from the first spectropolarimetric survey of $\gamma$ Dor pulsators}

   \subtitle{}

   \author{J. Labadie-Bartz\inst{1,2}\fnmsep\thanks{E-mail: jmila@dtu.dk}
        \and R. Ouazzani \inst{2}
        \and C. Neiner \inst{2}
        \and V. Antoci  \inst{1}
        \and P. Stanley \inst{3}
        \and T. Natan \inst{3}
        \and K. Thomson-Paressant\inst{2, 4}
        \and S. D. Chojnowski \inst{5}
        \and B. Mas Sanz \inst{1}
        \and O. Dürfeldt Pedros \inst{1}
        \and V. Petit \inst{3}
        }

   \institute{DTU Space, Technical University of Denmark, Elektrovej 327, Kgs. Lyngby 2800, Denmark 
   \and
   LIRA, Observatoire de Paris, Universit\'e PSL, CNRS, Sorbonne Universit\'e, Universit\'e Paris Cit\'e, CY Cergy Paris Universit\'e, 92190 Meudon, France
   \and
   Department of Physics and Astronomy, Bartol Research Institute, University of Delaware, 19716, Newark, DE, USA
   \and
   School of Mathematics, Statistics and Physics, Newcastle University, Newcastle upon Tyne, NE1 7RU, UK 
   \and 
   NASA Ames Research Center, Moffett Field, CA 94035, USA
 }

   \date{Accepted July 22, 2026}

  \abstract
   {Magnetic fields can have an important influence on stellar structure and evolution. In intermediate-mass (A- and F-type) stars, there are many known stars with directly measured strong, globally organized magnetic fields, as well as indications of weak and/or small-scale variable fields. However, among the intermediate-mass $\gamma$ Dor pulsators, there are no known stars with strong surface magnetic fields.}
   {The broad goal of this work is to search for evidence of strong, globally organized fields at the surface of $\gamma$ Dor pulsators.} 
   {We identified objects consistent with being $\gamma$ Dor pulsators based on an analysis of space photometry from the Transiting Exoplanet Survey Satellite (TESS) mission. A spectropolarimetric survey was then conducted on a subset of 47 of these objects, with a precision sufficient to detect dipolar surface magnetic fields down to a threshold of about 10 -- 100 G. }
   {We detected strong magnetism in three targets. However, upon closer inspection, none of these appear to be genuine $\gamma$ Dor pulsators. We found no evidence of surface magnetism in any of the remaining 44 objects. }
   {We conclude that either strong, globally organized magnetic fields and $\gamma$ Dor pulsation are mutually exclusive, or that such stars are exceedingly rare. A possible explanation is that strong global fields inhibit the excitation mechanism, which prevents $\gamma$ Dor pulsations from being driven in strongly magnetic intermediate-mass stars. The dipolar surface magnetic field strength upper limits we derive for this sample ($\lesssim$ 50 -- 100 G) provide valuable constraints for surface boundary conditions for asteroseismic models that include magnetism for $\gamma$ Dor stars. }


   \maketitle
   \nolinenumbers

\section{Introduction} \label{sec:intro}

The magnetic fields in stars are one of the major stumbling blocks of stellar evolution models, and yet it is ubiquitous in the Hertzsprung–Russell diagram (HRD). Magnetism affects stars at all evolution stages from star-forming molecular clouds \citep{McKee2007} to cooling white dwarfs \citep{Ferrario2020}.  

The space photometry revolution, with CoRoT and particularly {\it Kepler}, has allowed us to measure internal rotation rates in stars through the asteroseismology of mixed modes \citep{2010ApJ...713L.176B, 2012Natur.481...55B, 2012ApJ...756...19D, 2012A&A...544L...4E, Mosser2012} and g modes \citep{VanReeth2015,Christophe2018, Li2022}. These measurements have proven that evolutionary models that mainly include hydrodynamical transport processes for angular momentum and chemicals in radiative regions \citep{Zahn1992,Chaboyer1992} failed to explain the slowly rotating cores of intermediate-mass stars, measured from the main sequence \citep[MS;][]{Ouazzani2019,2019ARA&A..57...35A,2019MNRAS.485.3248M} to the tip of the Red Giant Branch \citep[RGB; see, e.g.,][]{Marques2013}. Remaining within the hydrodynamic framework, transport by internal gravity waves \citep{Talon&Charbonnel2005, Pincon2017} and oscillation modes \citep{Belkacem2015} has been explored, but this fails to counterbalance the evolution of angular momentum driven by structural changes in the right evolution stages. This suggests that purely hydrodynamical transport processes cannot explain the observations of the Sun, subgiant and red giant stars, or intermediate-mass MS stars.

Magnetism has been considered as one of the key ingredients in stellar interiors that can facilitate the redistribution of angular momentum \citep[e.g.,][]{2005A&A...440L...9E,Cantiello2014,2023A&A...677A...6M}. Indeed, stellar interiors are expected to be magnetized, and even weak magnetic fields combined with differential rotation can lead to efficient angular-momentum transport through Maxwell stresses \citep{Mestel&Weiss1987}. \cite{Spruit2002} proposed a self-sustained dynamo mechanism based on the Tayler instability \citep{Tayler1973}, operating in radiative zones. \cite{Fuller2019} later introduced a revised prescription, primarily differing in the saturation of the Tayler instability, leading to stronger magnetic fields and enhanced angular-momentum transport. However, neither formulation is able to simultaneously reproduce the rotation profiles of subgiants and red giants \citep{Cantiello2014, Eggenberger2019}. While early MHD simulations did not capture this dynamo process \citep{Zahn2007}, recent numerical studies have identified Tayler-like instabilities consistent with both the \cite{Spruit2002} and \cite{Fuller2019} prescriptions over a broader range of parameters \citep{Petitdemange2023, Petitdemange2024}. These studies provide physically motivated prescriptions that can be incorporated into stellar evolution codes \citep{Manchon2025}. Ultimately, only direct comparisons with asteroseismic constraints will allow the role of magnetic fields---for example through the Tayler–Spruit instability or core dynamo processes---in the transport of angular momentum within stellar radiative zones to be reliably assessed.

Asteroseismic inferences and measurements of internal magnetism have been especially notable in intermediate-mass red giant stars through the analysis of their ubiquitous solar-like oscillations. These include the suppression of dipolar mixed modes \citep{2016Natur.529..364S,Muller2025} and observed frequency shifts \citep[][and following works]{Li2022,Deheuvels2023}. Surface magnetic fields in red giants have also been directly detected through spectropolarimetry \citep{Auriere2015}.

Asteroseismic studies of internal magnetism in MS intermediate-mass stars have proven more challenging relative to their more evolved counterparts for several reasons, including more rapid rotation in MS stars and difficulties in mode identification (which is more straightforward for solar-like oscillations). However, progress has been made in describing the predicted effects of strong internal magnetism on low-frequency pulsation \citep[e.g.,][]{2019A&A...627A..64P, 2020A&A...638A.149V,Dhouib2022, Lignieres2024}. These theoretical developments have recently been applied to observations from Kepler, leading to the first asteroseismic detection of internal magnetism in an intermediate-mass MS star in \citet{Takata2026}. Several studies focusing on rotation have demonstrated that intermediate-mass MS stars rotate near rigidly, so there is already efficient angular-momentum transfer while on the MS \citep[e.g.,][]{2014MNRAS.444..102K, 2018A&A...618A..24V, 2019ARA&A..57...35A, 2019MNRAS.485.3248M, 2020MNRAS.491.3586L}. 
The lack of detectable pulsations can likewise constrain internal magnetic-field strengths in g-mode stars. When the core field exceeds a critical threshold, gravity modes are expected to convert into Alfvén waves and become strongly damped, effectively suppressing observable oscillations \citep{Fuller2015}. This principle has been used to set upper limits on internal fields in a slowly pulsating B (SPB) star \citep{Lecoanet2022} and is now being applied to $\gamma$ Dor stars \citep{2026A&A...711A.197D}.

\setcitestyle{notesep={; },round,aysep={},yysep={;}} 

Numerous spectropolarimetric observing campaigns have been carried out to describe the incidence rates and properties of surface magnetic fields for stars on the upper MS. In hot OB- and early A-type stars, about 10\% host strong, globally organized, and apparently stable magnetic fields \citep{2014IAUS..302..265W, 2006A&A...450..777B}. These are generally understood as fossil fields from an earlier evolutionary stage (e.g., the concentration of a magnetic field from a collapsing molecular cloud, or field generation during pre-MS evolution or through a merger event). The incidence of magnetically chemically peculiar (mCP) stars seems to drop quickly with decreasing mass for (near) MS stars with spectral types between about mid-A to mid-F, suggesting that strong global fields are rarer for the lower mass population of intermediate-mass stars \citep{2007pms..conf...89P,2019MNRAS.483.2300S}. Nevertheless, magnetic incidence rates in mid-A to mid-F stars are relatively poorly studied, and there are several confirmed cases of strong magnetism in this mass regime \citep[e.g.,][see also Sect.~\ref{sec:discussion_strongly_magnetic}]{1961Natur.189..739P, 2003A&A...404..669K, 2025A&A...704A.134T}. 

Spectropolarimetric detections of surface magnetism are not limited to strong fossil fields. At least some intermediate-mass stars have ultra-weak global fields with strengths on the order of a few Gauss or lower \citep[][see also Sect.~\ref{sec:weak_global_fields}]{2009A&A...500L..41L, 2010A&A...523A..41P, 2011A&A...532L..13P, 2022A&A...666A..20P, 2016MNRAS.459L..81B, 2016A&A...586A..97B, 2017MNRAS.468L..46N}. Detecting such weak fields is prohibitively expensive and practically impossible for all but the most optimal targets, which are exceptionally bright and have a low projected rotational velocity \citep[e.g.,][]{2016A&A...586A..97B, 2016MNRAS.459L..81B}. Weak magnetic fields (on the order of 1 G) have also been observed in mid- and late-F-type stars and are hypothesized to be of dynamo origin \citep[][see also Sect.~\ref{sec:dynamo_fields}]{2020MNRAS.494.5682S}. Photometric variability offers strong indirect evidence of dynamo-driven magnetism in A- and F-type stars. Many objects such as the “hump-and-spike” stars \citep{2013MNRAS.431.2240B,2018MNRAS.474.2774S,2023MNRAS.520..216H, 2023MNRAS.524.4196H} display rotational modulation and time-dependent surface structures  consistent with starspot activity \citep{2025A&A...696A.111A}. The associated fields are likely weak and multipolar, potentially placing them below the detection thresholds of current spectropolarimetric diagnostics.

Pulsating stars with directly measured surface magnetic properties are especially valuable for gaining insight into the influence of magnetism throughout the star. Indeed, nearly all classes of classical pulsators on the upper MS have members that are now confirmed to host strong magnetic fields. These include the $\beta$ Cephei stars \citep[e.g.,][]{2012A&A...537A.148N, 2013A&A...555A..46H}, SPB stars \citep[e.g.,][]{2013A&A...557L..16B}, and $\delta$ Scuti stars \citep[e.g.,][]{2015MNRAS.454L..86N,2025A&A...704A.134T}. The rapidly oscillating Ap (roAp) stars are all magnetic, of intermediate mass, and pulsate in high-frequency p modes \citep{1990ARA&A..28..607K,2019MNRAS.487.3523C,2021MNRAS.506.1073H,2024MNRAS.527.9548H}. The classical Be stars, which pulsate as a class \citep{2003A&A...411..229R, JLB2022}, are one exception where none are known to be strongly magnetic following a systematic spectropolarimetric observing campaign as part of the MiMeS survey \citep{Wade2016}. This is not unexpected, as strong magnetism is incompatible with the circumstellar disks of Be stars \citep{2018MNRAS.478.3049U}. The $\gamma$ Dor pulsators are the other exception---no surface magnetic fields have been directly detected to date.

$\gamma$ Dor stars are intermediate-mass ($\sim$1.3 -- 2 M$_{\odot}$), near-MS stars with low-frequency, non-radial pulsation \citep{1994MNRAS.270..905B, 1999PASP..111..840K}. The gravity mode pulsations in $\gamma$ Dor stars are of high radial order and are driven by convective blocking in a thin outer convective zone \citep{2005A&A...435..927D}. The gravity mode pulsations in $\gamma$ Dor stars are most sensitive to the near-core regions and are thus important probes of interior conditions. 

We set out to perform a spectropolarimetric survey of $\gamma$ Dor stars with the aim of detecting strong global magnetic fields or providing upper limits in the case of non-detections. The target selection and observing strategy is described in Sect.~\ref{sec:sample}. An analysis of space photometry and the spectroscopic and spectropolarimetric data for the strongly magnetic objects and the non-detections and upper limits are presented in Sect.~\ref{sec:analysis}. Section~\ref{sec:discussion} discusses our findings and attempts to contextualize them in relation to the overall picture of magnetism in intermediate-mass stars. Conclusions are given in Sect.~\ref{sec:conclusions}.

\section{Sample and data} \label{sec:sample}

With the growing number of pulsating stars on the upper MS with confirmed magnetic fields, a main driver behind this project was to attempt to discover a $\gamma$ Dor pulsator with a strong magnetic field. Our strategy was to use space photometry, primarily from TESS, to identify A- and F-type stars with signals consistent with $\gamma$ Dor pulsation, and ideally with hints of rotational modulation. For 47 of these stars, we acquired spectropolarimetric data with the Echelle SpectroPolarimetric Device for the Observation of Stars \citep[ESPaDOnS;][]{2006ASPC..358..362D}, operating on the Canada France Hawaii Telescope (CFHT) at Mauna Kea Observatory in Hawaii. 

Our target selection is biased and is not representative of the $\gamma$ Dor population as a whole. Targets were selected to maximize the likelihood of having a strong magnetic field while remaining feasible for spectropolarimetric observations, which favors bright, slowly rotating stars. Considering that many $\gamma$ Dor stars are found outside of the standard instability strip, no strict limits were imposed on stellar parameters, although most targets lie in or near the $\gamma$ Dor instability strip in the HRD. The only firm requirements were the low-frequency variability consistent with $\gamma$ Dor pulsation (i.e., multiple signals at frequencies $\lesssim$ 5 d$^{-1}$) and a spectral type in the A–F range. From the stellar parameters and uncertainties determined in Sect.~\ref{sec:stellar_parameters}, 38 of the 47 stars lie within or near the standard $\gamma$ Dor instability strip from \citet{2005A&A...435..927D} (two are too cool, three are too evolved, and six are too hot).

Priority was given to stars showing rotational modulation, often considered an indirect proxy for magnetism \citep[e.g.,][]{2019MNRAS.487..304D, 2023A&A...676A..55L, 2024A&A...689A.208T}, although such signatures proved difficult to identify in this sample (see Sect.~\ref{sec:tessphot}). Targets with long space-based photometric time series were also preferred. In practice, the integration time required to reach the desired signal-to-noise ratio (S/N) was the dominant constraint, biasing the sample toward brighter stars with lower $v \sin i$.

Our target spectropolarimetric S/N threshold was chosen to achieve sensitivity to global dipolar magnetic fields with a strength of 100 G. Nearly all high- and intermediate-mass stars with stable global magnetic fields have strengths of 300 G or more \citep[e.g.,][]{2007A&A...475.1053A, 2019MNRAS.490..274S}, although there are some exceptions with dipolar field strengths down to about 70 G \citep[e.g.,][]{2016A&A...589A..47A, 2019A&A...621A..47K, 2023MNRAS.521.3480K}. However, the population of these stars with stable global magnetic fields typically have spectral types of mid-A and earlier (however, see also Sect.~\ref{sec:discussion_strongly_magnetic}). The magnetic properties of late-A and mid- or early-F stars as a whole are poorly understood. We therefore chose a lower magnetic detection threshold to be sensitive to weaker fields. The spectroscopic S/N required for detecting a 100 G field in Stokes V was calculated considering brightness, $v \sin i$, and spectral type (which dictates the S/N gain from incorporating the LSD technique; generally, cooler stars exhibit more spectral lines and thus a higher LSD gain). Exposure times were kept short relative to the highest frequency photometric signals so that the line profiles would not vary appreciably over a spectropolarimetric sequence ($t_{\rm sequence} < 0.05 \times P_{\rm pulsation}$). However, some exceptions had to be made, especially for stars with very high frequency photometric signals (e.g., HD 102590, TYC 2430-1205-1, and HD 184875). For these exceptions, we do not expect that pulsational line-profile variations (LPVs) negatively impact our detection versus non-detection results, but they would impact the value of the field strength derived from the spectropolarimetric data in the case of a field detection.

\section{Analysis and results} \label{sec:analysis}

\subsection{TESS photometry} \label{sec:tessphot}

The Transiting Exoplanet Survey Satellite \citep[TESS;][]{2015JATIS...1a4003R} has provided high-precision space photometry for nearly the entire sky.
We used the TESS full frame images (FFIs) to extract and analyze photometry for our targets. An important part of this analysis was to check for blends to ensure that all detected signals originated on-target and not from a nearby (projected onto the sky) neighbor. We confirmed that all signals do indeed originate on-target by inspecting the frequency content of each pixel in the vicinity of the target and considering nearby \textit{Gaia} sources, as in \citet{2023A&A...676A..55L}, for example. All of our targets have several signals in the low-frequency regime ($\lesssim$5 d$^{-1}$), consistent with the expectations for $\gamma$ Dor stars.

The frequency spectra calculated from TESS photometry are shown in figures in Appendix~\ref{apx:LSD} for all stars in our sample. Most targets had between one and three consecutive TESS sectors. The longest set of consecutive TESS sectors was used to make the frequency spectra plotted in Appendix~\ref{apx:LSD}, rather than using the entire set of TESS data to avoid complex window functions. 
Besides the low-frequency signals typical of $\gamma$ Dor pulsators (multiple signals, often roughly organized in groups, and below about 5 d$^{-1}$), six of the targets have additional signals at higher frequencies. These are briefly discussed in Appendix~\ref{sec:other_pulsation}.

Since the majority of stars with surface-detected magnetism exhibit rotational modulation \citep[e.g.,][]{2019MNRAS.487..304D}, ideal magnetic $\gamma$ Dor candidates should exhibit both rotational modulation and pulsation. However, it proved difficult to find this combination of signals for several reasons. 
The range of frequencies in which $\gamma$ Dor pulsations are observed overlaps with the range of possible rotation frequencies ($\lesssim$3 d$^{-1}$), leading to ambiguities in identifying a rotation signal.
Due to spectropolarimetric observing constraints, stars with low $v \sin i$ dominate the sample where rotation periods may be long and poorly sampled in the short TESS sectors. In some cases, the photometric amplitude of rotational modulation can be very low, requiring Kepler-like precision such that the noise level in TESS data becomes problematic \citep[e.g.,][]{2020MNRAS.491.3586L}. 
Three of our targets show unambiguous rotational modulation that appears stable over the TESS observations: $\iota$ Phe (TIC 144276313; Sect.~\ref{sec:mag_iotphe}), TYC 2430-1205-1 (TIC 172414656; Sect.~\ref{sec:mag_TYC}), and HD 184875 (TIC 270610122; Appendix~\ref{sec:hd184875}). Three additional targets likely show time-variable rotational modulation: 78 UMa (TIC 229534764; Sect.~\ref{sec:mag_78UMa}), HD 196195 (TIC 276400050; Sect.~\ref{sec:dynamo_fields}), and I Pup (TIC 134497068; Sect.~\ref{sec:dynamo_fields}). Upon closer inspection of the TESS photometry after acquiring the spectropolarimetric data, BD+07 442 (TIC 387515681) shows only rotational modulation and is thus not included in the sample but is remarked upon in Appendix~\ref{sec:mag_bd+07442} due to a strong magnetic detection.

\subsection{Stellar parameters from SED and spectral fitting} \label{sec:stellar_parameters}

Preliminary stellar parameters ($T_{\rm eff}$, $\log g$, $v \sin i$) were required to plan our spectropolarimetric observations. These were collected from Simbad\footnote{\url{https://simbad.u-strasbg.fr/simbad/}}. When $T_{\rm eff}$ or $\log g$ values were not available, estimates were made based on the spectral type. For stars lacking $v \sin i$ estimates, an arbitrary value of 100 km s$^{-1}$ was chosen for calculating the required S/N needed to achieve our 100 G threshold.

After acquiring our spectropolarimetric data, the intensity spectrum was used to determine stellar parameters in a consistent way for the entire sample. 
The PySME\footnote{\url{https://pysme-astro.readthedocs.io/en/latest/index.html}} package is a Python implementation of the Spectroscopy Made Easy software developed to, among other things, fit grids of synthetic spectra to observed data to determine stellar parameters \citep{1996A&AS..118..595V,2023A&A...671A.171W}. We used PySME to fit our observed intensity spectra for each star (using the highest S/N spectrum whenever multiple observations were available), using the preliminary parameters as initial guesses. The line-by-line (LL) atmosphere model grids were used \citep{2004A&A...428..993S}, which cover a range of atmospheric parameters appropriate for our sample. PySME requires a line list, which was generated from the VALD database\footnote{\url{https://vald.astro.uu.se/}} \citep{1999A&AS..138..119K, 2000BaltA...9..590K, 2015PhyS...90e4005R} according to the preliminary value of $T_{\rm eff}$, using a $\log g$ of 4.0, a microturbulent velocity of 2 km s$^{-1}$,  a detection threshold of 0.01, and a solar chemical mixture.

Each observed spectrum was split into 18 sections from about 4000--6800 $\AA$ (avoiding H lines and regions heavily contaminated with tellurics), and fit with PySME to determine $T_{\rm eff}$, $\log g$, $v \sin i$, and the radial velocity (RV). As a first step, $\log g$ was fixed to 4.0, and $[M/H]$ to 0. $T_{\rm eff}$, $v \sin i$, and RV were then free parameters that were determined from fitting each spectral region. The mean value and standard deviations of these parameters were then adopted as our parameters and uncertainties, given in Table~\ref{tab:results}. Following this, to determine $\log g$ we used a narrow spectral region around the \ion{Mg}{I} triplet at 5167 $\AA$, 5173 $\AA$, 5184 $\AA,$ which are lines sensitive to pressure and therefore surface gravity in F-type stars \citep[e.g.,][]{1997A&A...323..909F}. Fixing $v \sin i$ and RV, we chose ten equally spaced values for $T_{\rm eff}$ within the lower and upper bounds determined from the previous step, keeping $\log g$ as the only free parameter. The final value and uncertainty for $\log g$ were taken as the mean and the standard deviation of $\log g$ for these ten iterations, respectively.

We note that our spectroscopically derived parameters are likely inaccurate for the small number of chemically peculiar stars ($\iota$ Phe, TYC 2430-1205-1, and possibly BD+07442 and 78 UMa), including the Am stars in our sample as we did not attempt a detailed abundance analysis or a fit to metallicity. Likewise, for SB2 systems, our parameters generally correspond to those of the primary star for higher contrast systems, which in some cases are contaminated by blended lines from a secondary. Among the obvious SB2 systems, only three potentially include stars of similar luminosity (V418 Peg, V1006 Cas, TYC 2430-1205-1). We did not attempt a detailed two-component spectral fitting. 

We additionally used the ``S-Phot visualizer sda''\footnote{\url{https://explore-platform.eu/sda/s-phot_visualizer_sda}} to perform spectral-energy-distribution (SED) fitting; this online tool is based on the PySSED\footnote{\url{https://github.com/iain-mcdonald/PySSED}} SED fitting software described in \citet{2024RASTI...3...89M}. The S-Phot tool considers parallax measurements and queries interstellar extinction maps and available broadband photometry to fit the SED and determine stellar parameters such as luminosity ($L$), temperature (T$_{\rm eff}$), and radius ($R$). 

While errors for a given SED fit are not provided by PySSED or S-Phot (due to impracticalities in constructing an appropriate error model), a comparison of PySSED results to ``gold-standard'' stellar parameters for Sirius A (A0) suggests that typical uncertainties in this temperature regime are $\sim$12\% in T$_{\rm eff}$, 8\% in $R$, and $\sim$35\% in $L$ \citep{2024RASTI...3...89M}.
Sirius A is an extreme case; all of the stars in our sample are (considerably) cooler and should be subject to smaller uncertainties. From a comparison of PySSED results to other high-confidence temperature determinations, typical errors in temperature over the range covered by our sample are $\sim$5\% in T$_{\rm eff}$ \citep{2024RASTI...3...89M}, but unfortunately such a comparison is not available for $L$ and $R$. We therefore adopted uncertainties of $\sim$5\% in T$_{\rm eff}$, 35\% in $L,$ and 8\% in $R$ (from Sirius). S-Phot is capable of fitting only a single atmospheric model for a given target, leading to inaccuracies for binaries.

Although we generally find good agreement between the stellar parameters determined from SED fitting and the spectroscopic analysis, a more detailed analysis of this sample would likely result in more accurate and precise stellar parameters. However, the magnetic analysis (which is the main aspect of this work) is not very sensitive to small inaccuracies in the stellar parameters as long as $T_{\rm eff}$ is approximately correct (Sect.~\ref{sec:specpol_analysis}). Representative examples of the spectral analysis and SED fitting are shown in Appendix~\ref{apx:spec_SED_fitting}.

\begin{figure}
    \centering
    \includegraphics[width=0.49\textwidth]{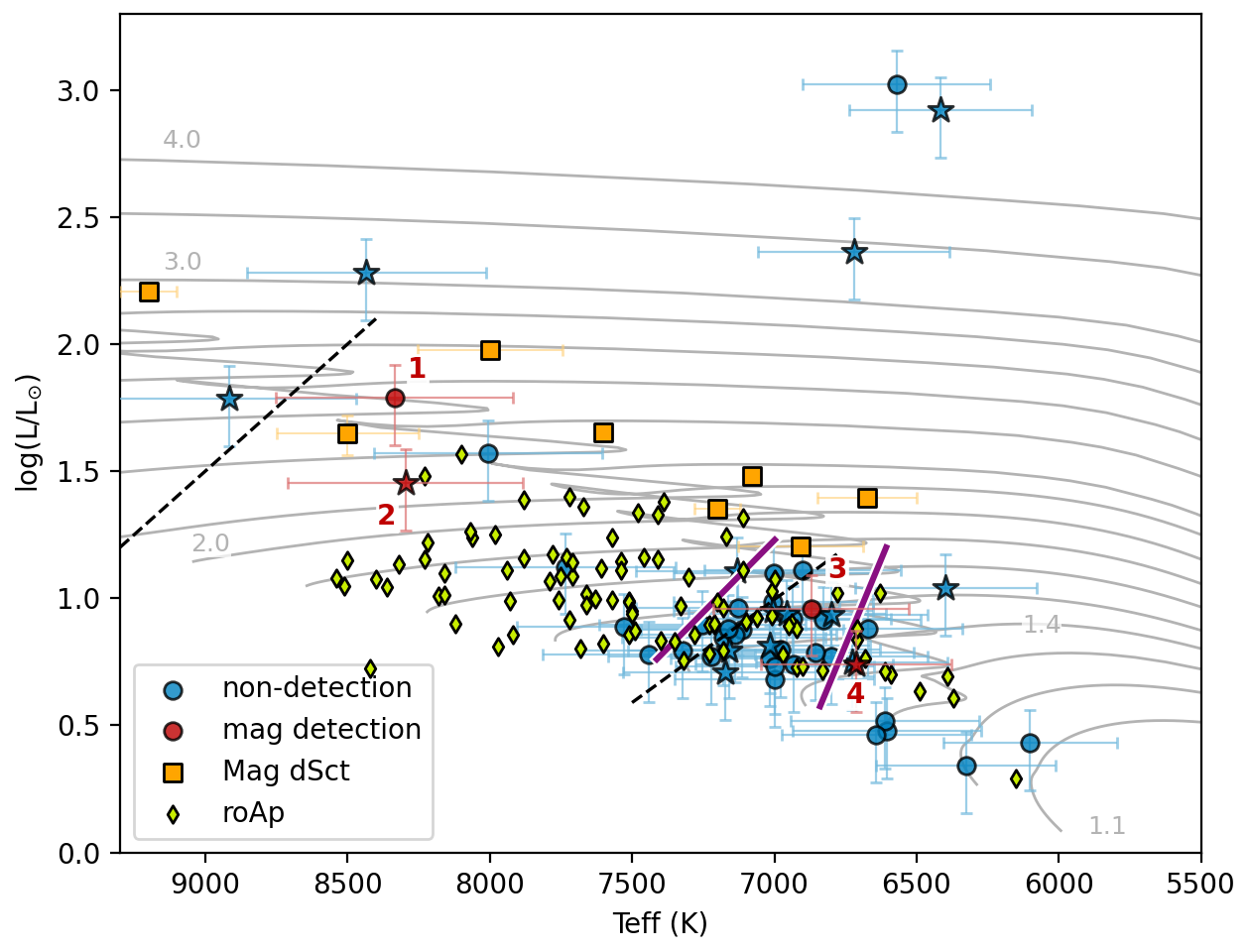}
    \caption{Luminosity and effective temperature as determined from SED fitting (Sect.~\ref{sec:stellar_parameters}) for the stars in our sample without magnetic detections (blue) and with a magnetic detection (red). Objects in our sample where we detect a magnetic field are  labeled 1 = $\iota$ Phe (Sect.~\ref{sec:mag_iotphe}), 2 = TYC 2430-1205-1 (Sect.~\ref{sec:mag_TYC}), 3 = BD+07 442 (Appendix ~\ref{sec:mag_bd+07442}), and 4 = 78 UMa (Sect.~\ref{sec:mag_78UMa}). The spectroscopic binaries are indicated by star symbols, while presumed single stars are marked by circles. Magnetic $\delta$ Scuti stars from \citet{2025A&A...704A.134T} are plotted as orange squares, and roAp stars from the lists of \citet{2021MNRAS.506.1073H,2024MNRAS.527.9548H} are indicated as yellow diamonds. The $\delta$ Scuti instability strip from \citet{2019MNRAS.485.2380M} is indicated by dashed black lines, and the $\gamma$ Dor instability strip is transposed from \citet{2005A&A...435..927D}. Luminosity (in log units) and effective temperature for our sample derived from SED fitting (Sect.~\ref{sec:stellar_parameters}). Evolutionary tracks are plotted with initial masses from 1.1 M$_{\odot}$ to 2.0 M$_{\odot}$, in steps of 0.1 M$_{\odot}$ from 2 to 3 M$_{\odot,}$ in steps of 0.2 M$_{\odot}$, and in steps of 0.5 M$_{\odot}$ from 3 to 4 M$_{\odot}$. }
    \label{fig:HRD}
\end{figure}

While it is valid to plot a spectroscopic HRD using T$_{\rm eff}$ and $\log g$ (a Kiel diagram), we preferred to use the luminosity and temperature from the SED fitting to plot our sample in an HRD as in Fig.~\ref{fig:HRD}. Especially with parallax information, we considered the SED fitting to provide more accurate values for luminosity compared to the spectroscopic determinations of $\log g$. The SED-derived parameters T$_{\rm eff}$, $L$, and $R$ are also given in Table.~\ref{tab:results}.

To guide the eye, evolutionary tracks are plotted in Fig. \ref{fig:HRD}. They were computed using the {\sc cesam2k20} code\footnote{freely available here: \url{https://www.ias.u-psud.fr/cesam2k20/}} \citep{Manchon2025} for masses between 1.1 and 4 M$_{\odot}$ at solar metallicity. The physics adopted for this grid of models is the same as in \cite{2025A&A...704A.134T}.

\subsection{Spectropolarimetric analysis} \label{sec:specpol_analysis}

Prior to any analysis steps, each spectrum was normalized with the normPlot\footnote{\url{https://github.com/folsomcp/normPlot/}} package. Each spectral order was normalized individually, the relatively noisy edges of overlapping orders were trimmed, and the orders were then merged. All following steps in the spectropolarimetric analysis were carried out using the Specpolflow\footnote{\url{https://github.com/folsomcp/specpolFlow/}} package \citep{2025JOSS...10.7891F}. As a sanity check, a few spectropolarimetric datasets were also processed with another set of routines that are commonly used  \citep[e.g., for the MiMeS survey;][]{2016MNRAS.456....2W}. All results were consistent within the error bars (and the uncertainties themselves were similar using the different analysis tools).

To analyze the spectropolarimetric data, we employed the least-squares deconvolution \citep[LSD;][]{1997MNRAS.291..658D, 2010A&A...524A...5K} method, which effectively combines information from many spectral lines to produce a high-S/N pseudo-averaged line profile. The end result for each observation is a profile of Stokes I (the intensity signal), a profile of Stokes V (the circular polarization signal), and a null profile, ``N'', which serves as a check against the circular polarization signal (the null profile should be flat if there are no problems with the data). The bin size in velocity was 1.8 km s$^{-1}$, the normalized line depth was set to 0.2, the normalized effective Lande factor was 1.2, and the normalized wavelength was 500 nm. The line centers were determined as the first moment of the Stokes I profiles over the velocity range $\pm v$ sin $i$ centered on the RV as determined from the spectral fitting routine described in Sect.~\ref{sec:stellar_parameters}. The integration range for the longitudinal field was set to be 1.2 times the $v \sin i$. The LSD profiles were not renormalized.

The LSD method requires a line list. For a given star, we used the same line list input as in the spectral analysis for determining the stellar parameters (Sect.~\ref{sec:stellar_parameters}), but with the effective temperature determined from the SED fitting rounded to the first two significant digits. A line mask was then generated after excluding the broad H-line regions and regions contaminated with tellurics and using a depth cutoff of 0.02. The individual line depths were then adjusted to match the observations \citep[as in, e.g.,][]{2017MNRAS.465.2432G}. 

The hemisphere-averaged line-of-sight component of the magnetic field, $B_{l}$, was then calculated from the resulting LSD profiles.
Following the example set by \citet{1992A&A...265..669D} and many subsequent works, a false-alarm-probability (FAP) algorithm was applied to each set of LSD profiles to determine the probability that a magnetic detection is legitimate. A definite detection (DD) is one where the \text{FAP} is $\lesssim 10^{-5}$, a marginal detection (MD) has $10^{-5} \lesssim \text{FAP} \lesssim 10^{-3}$, and a non-detection (ND) corresponds to $\text{FAP} \gtrsim 10^{-3}$.

The results of the magnetic analysis are presented in Sect.~\ref{sec:magnetic} for the magnetic detections, and the non-detections and upper limits for dipolar magnetic field strengths are described in Sect.~\ref{sec:nonmagnetic}. These results are summarized in Table~\ref{tab:results}, and plots for each star showing Stokes I, V, and N, as well as photometric signals from TESS are given in Appendix~\ref{apx:LSD}.
For visualization purposes, Stokes V and N profiles are plotted with the original data (points) and after smoothing via a sliding average with a window size of three data points (solid lines). All analyses were performed on the unsmoothed data.

\subsection{Binarity} \label{sec:binarity}

Spectroscopic binaries are revealed through either RV motion of the spectral features of one star (SB1), and/or through detection of two stellar components in an observed spectrum (SB2). Most of our sample was observed at only one epoch, but seven objects were observed twice, and one object was observed three times. Of these eight, five are definite binaries (HD 88815, PR Peg, V372 Peg, V418 Peg, and V1006 Cas; see figures in Appendix~\ref{apx:LSD}). This is a high incidence, so we expect that many of the other targets that were observed only once may also be binaries. 

Indeed, several additional objects observed by us only once are evidently SB2 systems based on their Stokes I profiles. These include HD 102590 (with a broad-lined primary, and two narrow-lined weaker components within the broad-lined profile), NY UMa (with a weak secondary that is resolved in velocity relative to the primary), TYC 2430-1205-1 (Sect.~\ref{sec:mag_TYC}), HD 2421 (with two narrow-lined components clearly separated in velocity), 78 UMa (Sect.~\ref{sec:mag_78UMa}), and V1012 Cas (similar to 78 UMa in Stokes I). 

The above binaries (listed as SB1 or SB2), as well as several additional candidates (listed as ``SB?'') are indicated in Table.~\ref{tab:results}. 
Many of our targets have distorted Stokes I profiles, but this is not necessarily due to binarity; pulsation can also induce line-profile asymmetries. With only one epoch for the majority of our targets, it can be difficult to distinguish between pulsation and binarity, as these two signals can be confused, especially if a genuine SB2 is composed of two (or more) stars with similar spectral type and $v$ sin $i$ for instance HD~184875 (Sect.~\ref{sec:hd184875}) and $\zeta$ Leo (Sect.~\ref{sec:zetLeo}). 

The lack of an obvious SB2 signal in Stokes I is not an indication that a given star is not a member of a binary. For example,  TYC 2430-1205-1 (Sect. \ref{sec:mag_TYC}) is potentially a multiple system based on RVs measured from archival Apache Point Observatory Galactic Evolution Experiment (APOGEE) spectra, yet this is not obvious from the Stokes I profile (Fig.~\ref{fig:LSD_TYC}).

\subsection{Stars with a detected magnetic field} \label{sec:magnetic}

Three stars in our sample exhibit a definite magnetic detection in Stokes $V$ (plus the non-pulsator BD+07 442; Appendix~\ref{sec:mag_bd+07442}). However, none of these seem to be magnetic $\gamma$ Dor pulsators. They are briefly discussed in the following three subsections. The first two, $\iota$ Phe and TYC 2430-1205-1, are $\delta$ Scuti pulsators. The remaining system, 78 UMa, is a binary with a magnetic signature only in the cool, narrow-lined, non-$\gamma$ Dor component.

\subsubsection{$\iota$ Phe (= HD 221760 = TIC 144276313)} \label{sec:mag_iotphe}

$\iota$ Phe is a chemically peculiar star (ApSrCrEu), and its magnetic field was first reported in \citet{2019MNRAS.483.3127S}, with a definite detection in four ESPaDOnS observations obtained in 2015. The $\langle B_{l} \rangle$ values varied between -72 $\pm$ 9 and 57 $\pm$ 9 G. However,  TESS photometry was not analyzed and a rotation period could not be established, preventing a reliable determination of the field geometry, nor was this star known to pulsate. 

In its TESS photometry, $\iota$ Phe exhibits high-frequency $\delta$ Scuti pulsation (out to $\sim$50 d$^{-1}$), low-frequency rotational modulation ($f_{\rm rot} = 0.3790$ d$^{-1}$), and additional non-rotational signals in the regime consistent with g modes (between about 1.5 -- 3 d$^{-1}$). This star was first noted as a possible $\gamma$ Dor and $\delta$ Scuti pulsator in \citet{2019MNRAS.487.3523C}. However, the notion of all non-rotational, low-frequency signals being combination frequencies is consistent due to nonlinear mode coupling \citep[e.g.,][]{1982AcA....32..147D, 2011MNRAS.414.1721B, 2018MNRAS.477.2183S, 2024A&A...687A.265V}. To check this, we searched for combinations using the $\delta$ Scuti frequencies between about 9 and 19 d$^{-1}$, treating these as potential parent modes. The five signals between 2 and 3 d$^{-1}$ correspond to simple differences from among the selected $\delta$ Scuti parent modes (i.e., $f_{i} - f_{j}$). The signals at $\sim$1.6 d$^{-1}$ and between 4 and 5 d$^{-1}$ match combinations of the form $nf_{i} - mf_{j}$, where $n$ and $m$ can take the values 1, 2, or 3. Therefore, these lower frequency signals may not be self-excited (e.g., via an opacity mechanism or convective blocking) modes, they may inherit their energy from the p modes. Relative to the rotation frequency, these non-rotational, lower frequency signals are higher than expected for high-radial-order g modes typical of $\gamma$ Dor pulsators (see also the discussion in Sect.~\ref{sec:mag_TYC}).

\begin{figure*}
    \centering
    \includegraphics[width=0.95\textwidth]{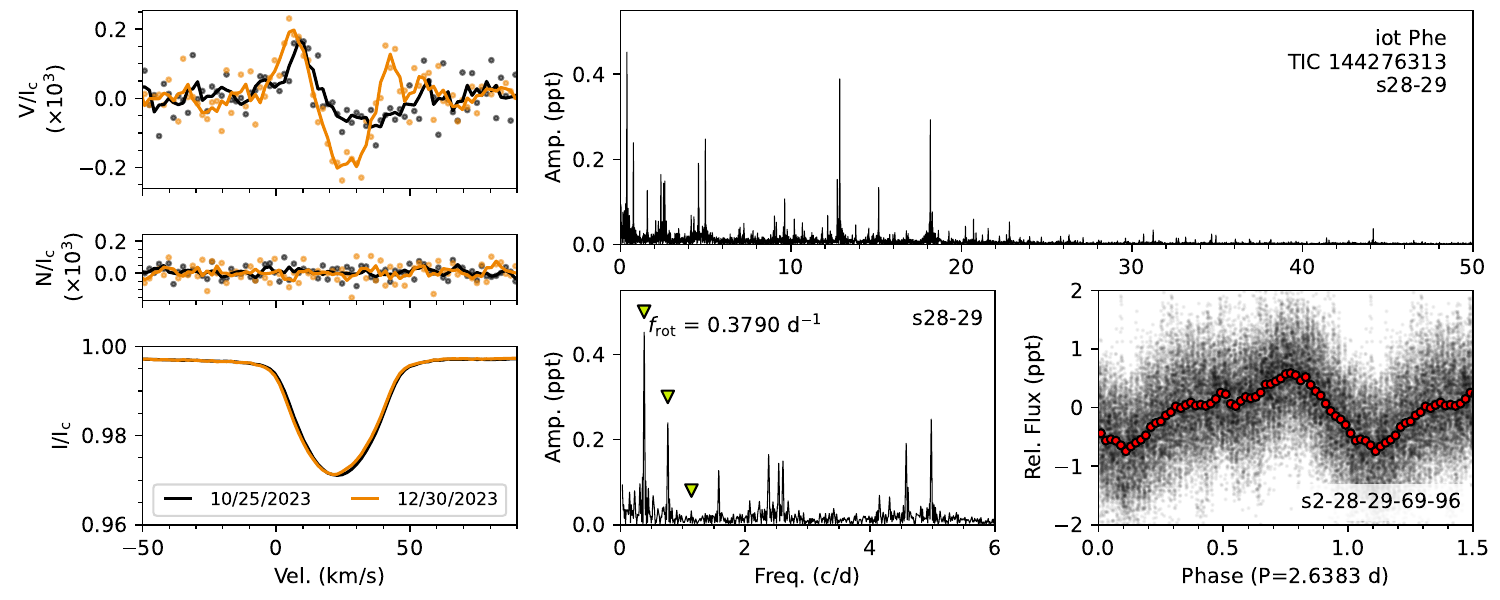}
    \caption{Data for $\iota$ Phe. \textit{Left:} Stokes $V$, $N$, and $I$ (from top to bottom) for the two spectropolarimetric observations. The solid lines for Stokes $V$ and $N$ are a sliding average with a bin size of three points. \textit{Right:} TESS frequency spectrum calculated from sectors 28 and 29, with the top panel extending to high frequencies and the lower panel emphasizing the low-frequency regime. The rotational frequency and its first two harmonics are indicated by yellow triangles.}
    \label{fig:LSD_iotphe}
\end{figure*}

In any case, $\iota$ Phe is an excellent magnetic target for more detailed asteroseismic analysis owing to the combination of high-frequency p modes, combination frequencies, and a precisely measured surface rotation frequency. Fig.~\ref{fig:LSD_iotphe} shows the Stokes V, N, and I profiles for our two new observations of $\iota$ Phe, as well as the signals from TESS. The small variations in Stokes I are also seen in the data presented in \citet{2019MNRAS.483.3127S} and are more consistent with surface features (rotating in and out of view) rather than RV motion due to binarity. From \textit{Gaia} DR3, the renormalized unit weight error (RUWE) has a value of 6.009, which indicates a high probability of photocenter motion due to an otherwise unseen binary companion (values of above 1.4 are generally considered as candidate astrometric binaries).

\subsubsection{TYC 2430-1205-1 (= TIC 172414656)} \label{sec:mag_TYC}

\begin{figure*}
    \centering
    \includegraphics[width=0.95\textwidth]{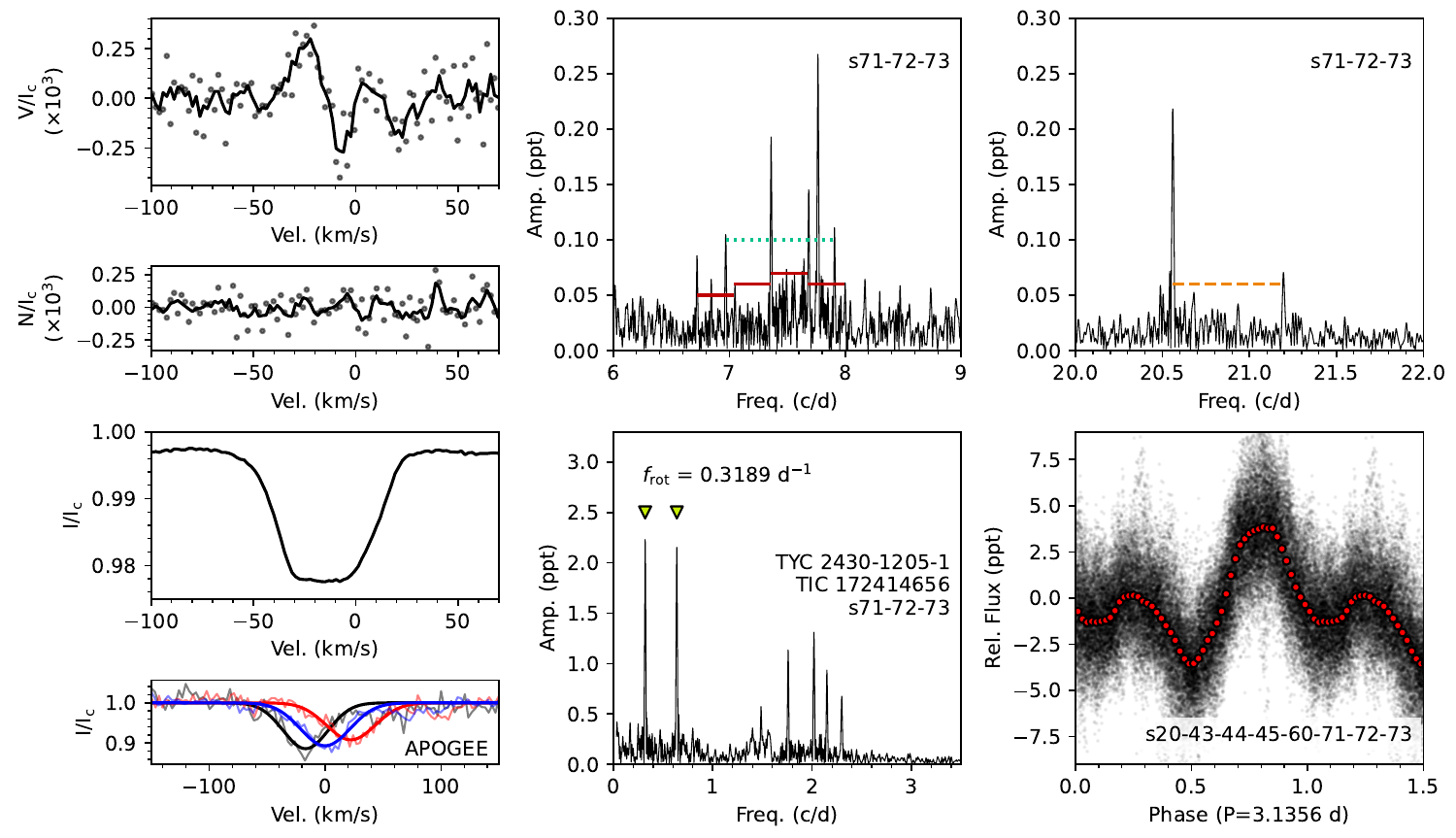}
    \caption{Similar to Fig.~\ref{fig:LSD_iotphe} but for TYC 2430-1205-1. The horizontal lines in the top middle and top right panels are at 1$\times f_{\rm rot}$ (solid red), 2$\times f_{\rm rot}$ (dashed orange), and 3$\times f_{\rm rot}$ (dotted green). The bottom right panel shows the entire TESS dataset phased to the rotation period, with 50 bins in phase (red dots), without removing any non-rotational signals. The bottom left panel plots three APOGEE observations of the \ion{Mg}{I}\,$\lambda$15770 line, each with a simple Gaussian fit (thicker curve) to illustrate the clear RV motion. }
    \label{fig:LSD_TYC}
\end{figure*}

The TESS frequency spectrum of TYC 2430-1205-1 shows rotational modulation, low-frequency signals consistent with g modes, and high-frequency p modes. Our one ESPaDOnS observation clearly indicates a magnetic field (Fig.~\ref{fig:LSD_TYC}). However, the RV of some of the lines vary in the nine available archival infrared APOGEE spectra \citep{2011AJ....142...72E, Blanton2017} from DR17 \citep{sdssDR17}.

The APOGEE spectra were taken over a 1124 day baseline and show lines of \ion{Ce}{III}, \ion{Ca}{II}, and \ion{Pr}{III}\,$\lambda$15679, corroborating the CP2 chemical peculiarity classification \citep[A5IV-VSrSiEu;][]{2022ApJS..259...63S}. These lines do not appear to shift in RV over the observed epochs. There is an additional set of lines in the spectrum that do exhibit RV variation, with a $max$ - $min$ variance of $\sim$50 km s$^{-1}$ (see bottom left panel of Fig.~\ref{fig:LSD_TYC}). 

One possibility is that this is a multiple stellar system. In this scenario, there may be three stars; the lines that show RV motion may belong to a star orbiting an unseen companion on a short orbit ($\sim$10 days), while the third star (with detected but non-moving lines) may be on a much wider orbit around this pair. A two-star system is less likely given the set of lines that do not move in RV (with typical uncertainties of $\pm$6 km s$^{-1}$). 

A second possibility is that there is only one star -- a magnetic star with strong chemical features on the surface. In this scenario, different ionic species can have distinct distributions over the stellar surface \citep[e.g.,][]{2022MNRAS.510.5821K, 2023MNRAS.521.3480K}. As the star rotates, some lines may appear to shift in RV, while others do not. With the relatively low S/N of the APOGEE data and only a single ESPaDOnS observation, it is not possible to compare how the finer structure of lines belonging to a given ion change with the rotation period. 

Either, or perhaps both, of the above scenarios can only be explored with more spectroscopic and/or spectropolarimetric data. Nevertheless, we can definitely conclude that TYC 2430-1205-1 contains a magnetic Ap star given the spectroscopic chemical peculiarities, the magnetic Stokes V signature, and the photometric rotational modulation.

Several of the higher frequency p modes exhibit frequency spacings very close to one, two, and three times the rotational frequency inferred from TESS (Fig.~\ref{fig:LSD_TYC}). 
These splittings are not perfectly exact (ranging within about $\pm$3\% of the rotation frequency of 0.3189 d$^{-1}$) and are thus consistent with rotational splitting \citep{2000ASPC..210..267G}.
We therefore presume that the magnetic Ap component hosts $\delta$ Scuti pulsation. 

The nature of the non-rotational, low-frequency signals is less clear. There are no simple linear combinations of the higher frequency signals that can produce the lower frequency signals as in $\iota$ Phe (Sect.~\ref{sec:mag_iotphe}). In this paragraph, we discuss whether or not it is plausible that these are high-radial-order g modes typical of $\gamma$ Dor stars in the magnetic component (ignoring binarity for now). Given the location of these signals relative to the rotation frequency, this seems unlikely. In \citet{2020MNRAS.491.3586L}, there are 39 $\gamma$ Dor pulsators with rotation frequencies between 0.2 and 0.5 d$^{-1}$ (i.e., similar to the rotation frequency of TYC 2430-1205-1); the majority are more rapidly rotating. Of these 39 slower rotators, the highest frequency observed from the $\ell = m = 1$ modes is on average 1.1 d$^{-1}$, which is significantly lower than the signals observed in TYC 2430-1205-1. It thus seems unlikely that a star rotating as slowly as TYC 2430-1205-1 can have $\ell = m = 1$ modes from $\sim$1.5 -- 2.4 d$^{-1}$. While modes of higher $m$ values are observed at higher frequencies, it is rare to observe, for example,  $\ell = m = 2$ modes in the absence of  $\ell = m = 1$ modes; only 2.8\% of the \citet{2020MNRAS.491.3586L} sample meet this criterion. Furthermore, the location of TYC 2430-1205-1 in the HRD is near the blue edge of the $\delta$ Scuti instability strip, far from the $\gamma$ Dor strip (Fig.~\ref{fig:HRD}), when its SED is fit as a single star. For these reasons, and considering the possibility of multiplicity, we cannot claim that the low-frequency signals are typical $\gamma$ Dor pulsations in the Ap star. 

If TYC 2430-1205-1 is a binary or triple system, it is possible the low frequency signals may be high-radial-order g modes from a non-Ap star, especially if rotating faster than the Ap star with rotation pushing the g modes to higher observed frequencies. The non-Ap component(s) may be closer or within the $\gamma$ Dor instability strip. Tidal effects may also be relevant. With the current data, the makeup of TYC 2430-1205-1 and the nature of the signals near 2 d$^{-1}$ remain unknown; additional spectroscopic and/or spectropolarimetric data are required.

\subsubsection{78 UMa (= HD 113139 = TIC 229534764)} \label{sec:mag_78UMa}

\begin{figure*}
    \centering
    \includegraphics[width=0.95\textwidth]{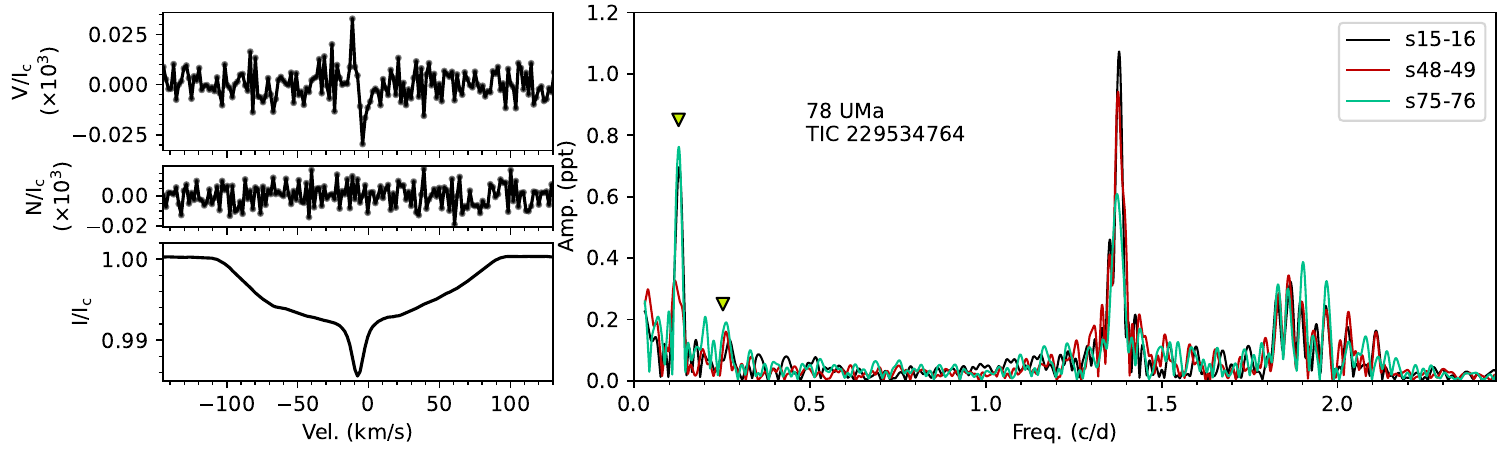}
    \caption{Similar to Fig.~\ref{fig:LSD_iotphe} for 78 UMa, an SB2 with a magnetic, narrow-lined star. A potential rotation frequency and its first harmonic for the narrow-lined star are marked by triangles. No smoothing was applied to the Stokes V and N profiles.}
    \label{fig:LSD_78UMa}
\end{figure*}

The Stokes I profile for 78 UMa clearly shows it is an SB2, with a broad-lined primary and a narrow-lined secondary (Fig.~\ref{fig:LSD_78UMa}). These two components are also clearly evident in the pure intensity spectrum, with the sharp lines being stronger at redder wavelengths (implying the narrow-lined star is relatively cool). The hotter primary has an effective temperature of $\sim$6700 K and a $v \sin i$ = 83 $\pm$5 km s$^{-1}$, and the secondary has $T_{\rm eff} \sim$5000 K and a $v \sin i$ of $\sim$5 km s$^{-1}$. While a detailed analysis of the stellar properties of both components was not performed, the secondary is certainly cooler than the sun and the composite spectrum is well fit with a synthetic spectrum for the aforementioned values of $T_{\rm eff}$ for both components; the secondary contributes $\sim$5\% of the flux in the spectral region around 5000 \AA ~\citep[which includes lines sensitive to temperature in this regime; e.g.,][]{2010A&A...512A..13S}. 

The position of 78 UMa in the HRD (Fig.~\ref{fig:HRD}) corresponds to the hotter, nonmagnetic, component. The position of the cool magnetic component in this binary system would be below both axes in Fig.~\ref{fig:HRD}.

A clear Stokes-V signal is evident, coinciding with the position and width of the narrow-lined component (Fig.~\ref{fig:LSD_78UMa}). The narrow-lined secondary is below the Kraft break and is therefore very likely to have a convective envelope and a dynamo magnetic field. Except for the velocity range spanned by the narrow-lined star, Stokes V is consistently flat across the profile of the broad-lined star. 

The many low-frequency signals in TESS (Fig.~\ref{fig:LSD_78UMa}) are clustered around 1.2 -- 2.2 d$^{-1}$ and 3.5 -- 4.5 d$^{-1}$;  these are most likely g modes of the more rapidly rotating broad-lined star, although the exact rotation frequency is unknown. There is also a peak in the TESS data at $f \approx 0.12$ d$^{-1}$ with one harmonic, which may be the rotation frequency of the cooler star. This candidate rotational signal varies significantly in amplitude between different TESS epochs. If this does trace rotational modulation of the narrow-lined magnetic star, it implies that the surface inhomogeneities vary significantly with time and are therefore unlikely to be caused by a stable fossil field, which is consistent with the cool temperature. Presuming both stars are of the same age, the cool star is relatively young in its MS evolution and may thus be expected to show a high level of activity. The strongest signal in TESS, near $f = 1.38$ d$^{-1}$ varies in amplitude, frequency, and shape from sector-to-sector; does not have any harmonics; and is consistent with unresolved pulsation signals (as opposed to rotational modulation of the broad-lined star). 

This system is thus composed of a rapidly rotating $\gamma$ Dor star without a strong magnetic field at its surface, and a much more slowly rotating and cooler star with a strong surface field probably of dynamo origin. The longitudinal magnetic field value measured from the LSD profiles including only the velocities corresponding to the narrow-lined star is $B_{z}$ = 7 $\pm$ 1 G. Of course, this is not an accurate value for the surface field of the secondary, as the observed flux is dominated by the hotter star. This example demonstrates that relatively weak and localized (in the sense that a Stokes V signature is present only in part of a line profile) magnetic fields are in principle detectable in our dataset. From the Bayesian analysis described in Sect.~\ref{sec:nonmagnetic}, the narrow-lined component was isolated, and presuming a dipolar topology the magnetic-field strength is between 26 and 149 G (at the 95\% credible regions) or between 35 and 76 G (at the 68\% credible regions).

\subsection{Magnetic non-detections and upper limits} \label{sec:nonmagnetic}

Besides the three objects described in Sect.~\ref{sec:magnetic}, an analysis of the remainder of our sample resulted only in magnetic non-detections. For these 44 objects with magnetic non-detections, we followed the Bayesian methods described in \citet{2012MNRAS.420..773P} to determine an upper limit on the dipole magnetic-field strength using the pyRaven\footnote{\url{https://veropetit.github.io/pyRaven}} package. 
First, the Stokes I profile of each star was fit with a Voigt profile convolved with a rotation kernel and then a Gaussian (for macroturbulence and spectral resolution) to extract $v \sin i$, the line strength parameter (log($\kappa$)), the macroturbulent velocity ($v_{mac}$), and the RV ($v_{rad}$). An average fit was obtained for stars with multiple observations (after correcting for $v_{rad}$). 
PyRaven generates synthetic Stokes-V profiles at all possible magnetic strengths and geometries and then compares them to the actual Stokes-V profiles of the objects to output a posterior probability density of the dipolar magnetic-field strength, B$_{pole}$; the inclination angle of the rotation axis ($i$); and the angle between $i$ and the magnetic dipole axis ($\beta$). The phase of each observation is unconstrained; the likelihood of each observation is marginalized over the possible phases before the probabilities are combined. 
We then calculated the upper bounds of the 95\% and 68\% credible regions and adopted these values as the dipolar magnetic-field upper limits.
A similar analysis was done for other spectropolarimetric datasets for samples without detected fields \citep[e.g.,][]{neiner2015b,2019MNRAS.489.5669P, 2023MNRAS.526.1728T}. 

The priors for the Bayesian analysis are as follows. For inclination, a flat sin$i$ prior between zero and 180 degrees, with a five-degree step size, was used to give equal weight to any orientation. For obliquity ($\beta$), a flat prior of between zero and 360 degrees with a five-degree step size was used. The phase also used a flat prior. The dipolar field strength, $B_{pole}$, used a modified Jefferys prior, which gives equal weight to every order of magnitude, with a flat prior for values below the Jefferys cutoff. This cutoff was chosen to be two times the grid step size, where we used 200 grid steps of between zero and three times the field strength for a pole-on view of the dipolar field that yields a Stokes-$V$ signal with the same amplitude as the noise level. A noise scaling parameter ($\sigma$ = $s/b$), where $s$ is the LSD error and $b$ is the scaling, was included to account for any additional noise from physical features that distort the LSD profiles (e.g., pulsation) with a grid between 0.1 and 2 with a step size of 0.1. The prior is the Jeffreys one for the noise scaling, and it is marginalized at the same time as the phase (i.e., each observation has an independent noise scaling). The key assumptions are that the star is well described by the weak field approximation (which is valid for this sample as our non-detections do not have very strong fields), the magnetic field is dipolar, and a Voigt profile with rotation and limb darkening and Gaussian broadening (for macroturbulence and spectral resolution) provides a reasonable fit to Stokes I.

\begin{figure}
    \centering
    \includegraphics[width=0.49\textwidth]{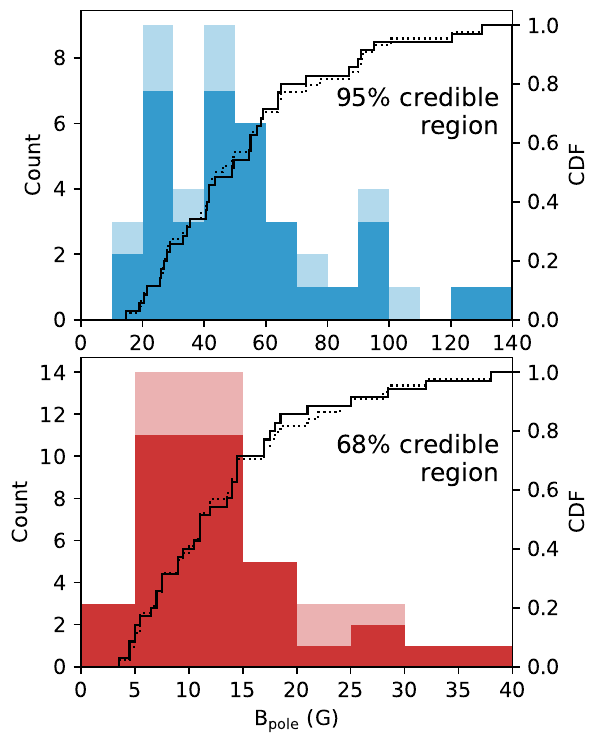}
    \caption{Histograms showing the dipolar magnetic-field strength upper limits for non-detections at the 95\% (upper) and 68\% (lower) credible regions. The darker shading is for the apparently single stars, and the lighter shading also includes the primary stars from the spectroscopic binaries. The solid black line in each panel is the normalized cumulative distribution function (CDF) for the apparently single stars, and the dotted line includes the primaries of binaries.  }
    \label{fig:upper_limits}
\end{figure}

Figure~\ref{fig:upper_limits} shows histograms of the dipolar field strength upper limits at the 95\% and 68\% credible regions for the 35 stars that appear as single in their Stokes I profile (excluding those with detected fields). At the 95\% (68\%) credible region, the mean value of the upper limits is 53 G (13 G). The primary components of the obvious spectroscopic binaries are added to these distributions with fainter colors in Fig.~\ref{fig:upper_limits}; this has little impact on the cumulative distribution function (plotted as a solid line for the apparently single stars only, and as a dotted line when including the primaries of the binaries). We presumed that the primaries (those with the dominant spectral features) are also the photometrically variable stars, but this is not guaranteed in all cases. These upper limits rule out the possibility of these stars having strong dipolar magnetic fields, as all such fossil fields have dipolar strengths on the order of hundreds of Gauss or more. It does not, however, rule out ultra-weak fossil fields (Sect.~\ref{sec:weak_global_fields}), dynamo-generated fields that emerge at the surface (Sect.~\ref{sec:dynamo_fields}), or internal magnetic fields (Sect.~\ref{sec:internalMag}). 

There are two objects with somewhat higher upper limits than the remainder of the sample. These are FO Cet, with an upper limit of 121 G (95\% confidence), and HD 184875, with an upper limit of 130 G (95\% confidence). For FO Cet, its $v \sin i$ as determined from our spectral fitting (59 km s$^{-1}$) is about 50\% higher than we assumed from its literature value. HD 184875 may be a binary with a similar-brightness companion, or its line profiles may be distorted by its unusual pulsation patterns (Appendix~\ref{sec:hd184875}). These factors led to an underestimation of the required exposure time to achieve a 100 G threshold for these two objects.

\section{Discussion} \label{sec:discussion}

We did not detect any evidence of a globally organized magnetic field in any of the  $\gamma$ Dor pulsators in our spectropolarimetric sample. This suggests that strong, globally organized magnetic fields, such as the fossil fields regularly observed in $\sim$10\% of OBA stars, are either intrinsically rare in $\gamma$ Dor stars, or that $\gamma$ Dor pulsation and strong global fields are mutually exclusive. Although three objects with a detected magnetic field at first glance appear to have signals consistent with $\gamma$ Dor pulsations, further scrutiny suggests the magnetic objects do not host high-radial-order g modes (Sects.~\ref{sec:mag_iotphe},~\ref{sec:mag_TYC}, and ~\ref{sec:mag_78UMa}).
However, our observational methods are not sensitive to internal magnetic fields or ultra-weak fossil fields, and they have limited sensitivity to small-scale (variable) surface fields. The remainder of this section discusses the implications of the lack of magnetic detections, draws comparisons to other relevant classes of stars, and explores methods complementary to spectropolarimetry for the study of magnetism in this stellar regime.

\subsection{Strongly magnetic stars overlapping with $\gamma$ Dors in the HRD} \label{sec:discussion_strongly_magnetic}

The majority of observational studies focusing on strong fossil fields involve O-, B-, and early A-type stars, which are far from the $\gamma$ Dor instability strip \citep[e.g.,][]{2016MNRAS.456....2W, 2017MNRAS.465.2432G, 2019MNRAS.490..274S, 2019MNRAS.483.3127S}. There are, however, several directly confirmed fossil fields in stars around and within the $\gamma$ Dor instability strip. Many of these cooler, intermediate-mass stars with strong magnetism are rapidly oscillating Ap (roAp) stars; this is a class of strongly magnetic stars with very high frequency pulsations, typically of $\sim$100 -- 300 d$^{-1}$ \citep{1990ARA&A..28..607K, 2019MNRAS.487.3523C}. For example, HD 213637 with T$_{\rm eff} \approx 6400$ K \citep{2003A&A...404..669K} and HD 101065 (Przybylski’s star, \citet{1961Natur.189..739P}) with T$_{\rm eff} \approx 6400$ K \citep{2010A&A...520A..88S} are both roAp stars with directly measured strong magnetism and temperatures consistent with mid-F spectral type. HD 24712, with T$_{\rm eff} \approx 7250$ K \citep{2009A&A...499..879S} is another roAp star near the blue edge of the $\gamma$ Dor instability strip. Other examples of strongly magnetic stars with fossil fields in or near the $\gamma$ Dor instability strip include HD 12932 \citep{2015A&A...583A.115B} and HD 51203 \citep{2019ApJ...873L...5C}. HD 41641, a $\delta$ Scuti pulsator (but not roAp), is yet another case \citep{2021MNRAS.500.1992T, 2025A&A...704A.134T}. This is not an exhaustive list, but it demonstrates that the region in the HRD occupied by $\gamma$ Dor pulsators is not a magnetic desert. None of the aforementioned stars are $\gamma$ Dor pulsators.

There are five roAp stars (either confirmed or candidate) presented in \citet{2021MNRAS.506.1073H, 2024MNRAS.527.9548H} for which the analysis of TESS photometry shows the presence of low-frequency variability consistent with g modes. However, three of these are binaries or higher order systems with similar-temperature or slightly cooler secondaries (TYC 9348-1029-1, HD 218439, and HD 76279). In these cases, \citeauthor{2021MNRAS.506.1073H} suspected that the low-frequency pulsational signals are not from the roAp star. The fourth, HD 10330, is only a candidate roAp star with marginally detected high-frequency roAp pulsation and no spectroscopic indication of chemical peculiarity nor a directly confirmed magnetic field. Analysis of the 20-second-cadence TESS data from sector 96 (August, 2025) does not show any significant peaks near the candidate frequencies ($\sim$236 d$^{-1}$) marginally detected in the earlier TESS data analyzed in \citet{2021MNRAS.506.1073H}; i.e., there is no convincing evidence that HD 10330 is a roAp star. The fifth, HD 193756, shows rotational modulation and signals consistent with g-mode pulsation (in addition to roAp pulsation). There does not seem to be direct evidence of binarity, neither is there strong evidence of the star being single. As pointed out in \citet{2024MNRAS.527.9548H}, the high Gaia RUWE of 3.5 may suggest the presence of a companion. Thus, there do not seem to be any confirmed roAp stars that also exhibit $\gamma$ Dor type pulsation, although HD 193756 is worth investigating further.

\subsection{Weak, apparently stable magnetic fields} \label{sec:weak_global_fields}

Ultra-weak magnetic fields, with strengths on the order of 1 G, have been directly detected via spectropolarimetry in a few intermediate-mass stars. 
The first such discovery was for the rapidly rotating star Vega (A0), which hosts a magnetic field detectable at its surface with a strength $\sim$2 orders of magnitude weaker than the strongly magnetic Ap stars. \citep{2009A&A...500L..41L, 2010A&A...523A..41P}. This field seems to be stable on timescales of years, but there are also hints of smaller scale variations that suggest the coexistence of this larger scale field and localized and variable dynamo fields \citep{2022A&A...666A..20P, 2025A&A...702A..19B}.

Weak magnetic fields have also been found in the chemically peculiar Am and Fm stars, which tend to be moderate to slow rotators. The first magnetic detection in an Am star was found for Sirius A with a disk-integrated line-of-sight field of $\sim$0.2 G \citep{2011A&A...532L..13P}. Similar field strengths, on the order of $\lesssim$1 G, were then found in the Am stars $\beta$ UMa and $\theta$ Leo \citep{2016A&A...586A..97B}. In all three cases, the integral of the Stokes-$V$ signature across the line profile is far from zero, in contrast to the signatures commonly observed in hot, strongly magnetic stars. 
The Am star Alhena A hosts a comparatively stronger field of $\sim$30 G and has a more traditional Stokes-$V$ profile \citep{2020MNRAS.492.5794B}, perhaps indicating that its magnetic field is of a different nature than the other Am stars. Am stars can host $\gamma$ Dor pulsation \citep[e.g.,][]{2005AJ....129.2026H,2024A&A...690A.104D}.

$\rho$ Pup is an intermediate-mass Fm star and $\delta$ Scuti pulsator with a weak $\sim$1 G apparently stable field \citep{2017MNRAS.468L..46N}. HD 73857 may be another similar magnetic $\delta$ Scuti pulsator, but it is not yet clear if its weak magnetic field is global and stable or variable and of dynamo origin \citep{2025A&A...704A.134T}. Both $\rho$ Pup and HD 73857 are in or near the $\gamma$ Dor instability strip.

Our spectropolarimetric data are of an insufficient S/N to detect fields at the $\lesssim$1 G level. Very few high- and intermediate-mass stars have been observed with such deep spectropolarimetric observations to achieve the required polarimetric precision to detect these weak fields. As pointed out in \citet{2009A&A...500L..41L}, it is possible that ultra-weak fields are common in intermediate-mass stars, but this has yet to be tested.

\subsection{Variable dynamo magnetic fields} \label{sec:dynamo_fields}

Dynamo-driven magnetic fields can produce variable star spots, causing a photometric signal at the rotation period that changes in shape with time as the spots evolve. While this behavior is ubiquitous in solar-like and lower mass stars, these signatures are far less common and found at much lower photometric amplitude, in stars with predominantly radiative envelopes. Nevertheless, some fraction of intermediate-mass stars exhibit this type of photometric variability. We list different types of intermediate-mass stars that seem to exhibit such variability and then discuss why this would be a plausible explanation for some stars in our sample.

A spectropolarimetric survey was performed on 53 stars with spectral types between F0 and F9 \citep{2020MNRAS.494.5682S}. Magnetic signatures were detected in 14 of these, with longitudinal magnetic-field strengths between 0.3 and 8.3 G. No fields were detected in stars with spectral types of F2 and earlier. Since most stars were observed only once, it is difficult to draw firm conclusions about the nature of these fields; i.e., whether they are weak fossil fields or dynamo generated. We examined TESS photometry for the 14 stars with detected magnetic fields and saw no evidence of $\gamma$ Dor pulsation in any of them. However, several showed signals consistent with variable rotational modulation. It is therefore possible, as pointed out in \citet{2020MNRAS.494.5682S}, that the detected magnetic fields in these mid- to late-type F stars are of dynamo origin.

The ``hump and spike'' stars are a class of intermediate-mass stars \citep{2018MNRAS.474.2774S} that show time-variable rotation \citep{2023MNRAS.520..216H, 2025A&A...696A.111A}. The rotational modulation manifests as a variable spike in the amplitude spectrum at the stellar rotation frequency (plus up to several harmonics). The favored explanation is that dynamo-driven magnetic fields cause the observed surface variability, although there is no direct spectropolarimetric confirmation yet \citep{2025A&A...696A.111A}. These stars exhibit Rossby pulsation modes as a rule (causing the hump located slightly below the rotation frequency in the amplitude spectrum), and some also exhibit signs of g-mode oscillations \citep{2023MNRAS.524.4196H}. Interestingly, the majority of the hump and spike stars are on the hotter side compared to the sample of \citet{2020MNRAS.494.5682S}, where outer convection zones are thought to be very thin.

Another sample of intermediate-mass stars that seem to exhibit variable rotational modulation are the $\delta$ Scuti stars analyzed in \citet{2023MNRAS.526.1728T}. Although no magnetic signature was detected in the spectropolarimetric data for any of these stars, the magnetic upper limits still allow for weak and/or dynamo fields. Weak magnetism has been detected in the $\delta$ Scuti star $\beta$ Cas, with evidence supporting a dynamo origin presented in \citet{2020A&A...643A.110Z}. 
As the hump and spike stars, most of these $\delta$ Scuti stars with potential rotational modulation are sufficiently hot such that stellar models predict only a very thin outer convection zone, which brings into question the mechanism for creating their purported surface spots.

Of the 611 $\gamma$ Dor stars observed with Kepler with clear period-spacing patterns  and analyzed in \citet{2020MNRAS.491.3586L}, about 10\% were found to show non-pulsational variability consistent with rotational modulation. An inspection of the candidate rotational signals in the Kepler data for these stars shows variations in the amplitude of the rotational signal and its harmonics, suggesting variable spots and thus dynamo-driven small-scale fields. These candidate rotation frequencies are bolstered by the fact that they are very similar (to within $\sim$5\%) to the near-core rotation rate inferred through the analysis of period-spacing patterns of g modes in \citet{2020MNRAS.491.3586L}. It is possible that an even higher fraction of this 611-star sample exhibits this type of rotational modulation, as the two $\gamma$ Dor stars analyzed in \citet{2026A&A...711A.197D} also show rotational modulation (one having a constant rotation signal, and the other a variable one), despite not being listed as such in \citet{2020MNRAS.491.3586L}. These stars are faint, and none have been observed with spectropolarimetry to our knowledge. 

\begin{figure*}
    \centering
    \includegraphics[width=0.95\textwidth]{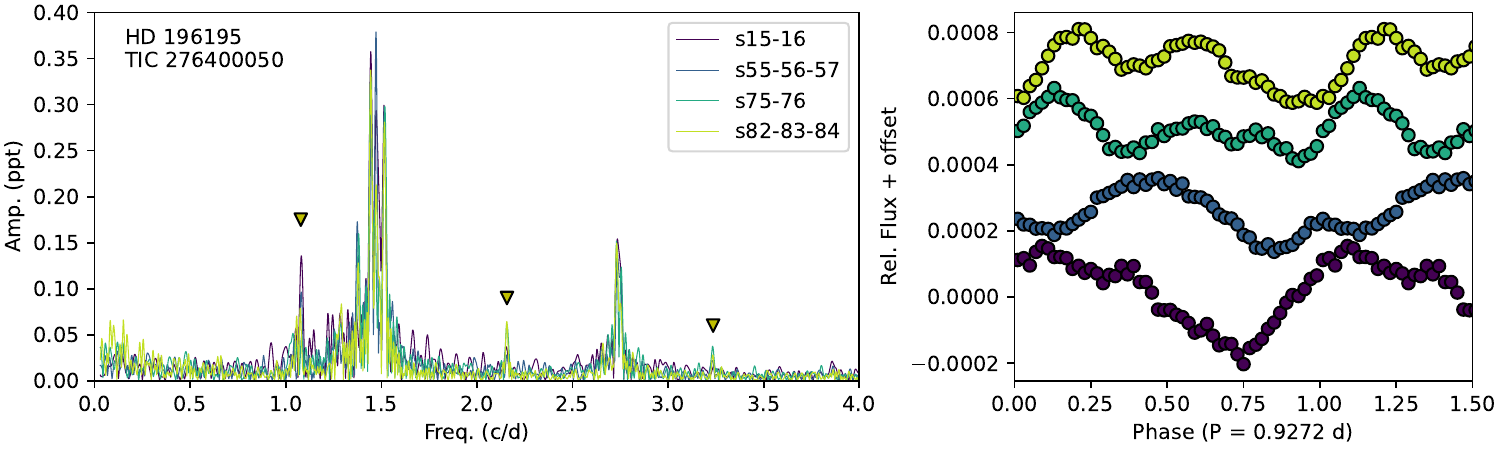}
    \caption{ \textit{Left:} TESS frequency spectra for four strings of photometry for HD 196195, with each color corresponding to a different set of consecutive TESS sectors as indicated in the legend. The candidate rotation frequency and its first two harmonics are marked by triangle symbols. \textit{Right:} Same four strings of photometry phased to the candidate rotation period, after removing the higher amplitude pulsational signals and binned in steps of 0.01 in phase. The same time zero point was used to phase these different data strings. 
    }
    \label{fig:rotmod_276400050_01}
\end{figure*}

Although only a small fraction of the genuine $\gamma$ Dor stars in our sample exhibit signals consistent with rotational modulation, there are a few practical problems with the detection of these signals (Sect.~\ref{sec:tessphot}). Nevertheless, the strong evidence of rotational modulation in about 60 $\gamma$ Dor stars in \citet{2020MNRAS.491.3586L}, and a few in the present work, support the notion that a fraction of $\gamma$ Dor stars have variable surface features that appear consistent with dynamo-generated magnetic fields. Figure~\ref{fig:rotmod_276400050_01} shows one such example for HD~196195 (= TIC 276400050), with a candidate rotational signal at 1.079 d$^{-1}$. i Pup (TIC 134497068) may also exhibit variable rotational modulation at $\sim$0.2 d$^{-1}$ with its most prominent g mode group near 1 d$^{-1}$. HD~184875 shows constant rotational modulation (see Sect.~\ref{sec:hd184875}), but without a magnetic detection.

\subsection{Internal magnetic fields in $\gamma$ Dor stars} \label{sec:internalMag}

The absence of direct magnetic field detections at the surface of $\gamma$ Dor stars does not exclude the presence of significant internal fields. 
Asteroseismology can indirectly probe internal magnetism, as the fields can cause frequency shifts and/or mode suppression (see Sect.~\ref{sec:intro}).

In a recent study, \citet{Takata2026} exploited observed frequency shifts to describe the internal structural and magnetic properties of the star KIC 9244992 (F0), a hybrid $\gamma$ Dor and $\delta$ Scuti pulsator, through analysis of asymmetric splitting of its g modes observed with Kepler photometry. The minimum radial and azimuthal magnetic field components within the inner 50\% in radius were determined to be B$_{\rm r} = 3.5 \pm 0.1$ kG and B$_{\rm \phi} = 92 \pm 7$ kG. This was the first such work using this technique on a MS star. KIC 9244992 is faint ($V_{\rm mag}$ = 14.2), and no spectropolarimetric data are available. However, their modeling does not require a detectable field at the surface to fit the data.

The ability of strong magnetic fields to suppress pulsation modes is useful in the study of internal magnetism in red giants (see Sect.~\ref{sec:intro}) and has been used for one slowly pulsating and strongly magnetic B star \citep[][see also Sect.~\ref{sec:mag_SPB}]{Lecoanet2022}. \citet{2026A&A...711A.197D} applied similar methods to \citet{Lecoanet2022} to two $\gamma$ Dor pulsators, finding upper limits for the near-core magnetic-field strength of roughly 10--100 kG. Assuming a current-free dipole field configuration extending all the way to the surface, a surface strength of roughly 5--90 G would be expected for their targets. Emergent surface fields on the higher end of this range should be detectable in at least half of our sample (Sect.~\ref{sec:nonmagnetic}). However, for magnetic fields generated by the core dynamo, these may not reach the stellar surface within the MS lifetime due to a relatively low magnetic buoyancy as discussed in \citet{1989MNRAS.236..629M}, \citet{2016ApJ...829...92A}, and \citet[][and references therein]{2024A&A...691A.326H}. Different field configurations and internal structures and processes can act to either increase or decrease the time required for a core-generated field to rise to the surface, so it is not yet clear what the expectations are \citep{1978A&A....68...57S, 2003ApJ...586..480M, 2004MNRAS.348..702M, 2024A&A...691A.326H}.

For stars with a strong fossil field, the near-core field strength can be significantly higher than in situations where internal magnetism is generated only via the core dynamo \citep[e.g.,][]{2009ApJ...705.1000F, Lecoanet2022}. This can lead to the suppression of g modes, in particular those of high radial order. However, g modes have been observed in stars with strong fossil fields (Sect.~\ref{sec:mag_SPB}), so mode suppression may not (solely) explain the lack of strongly magnetic $\gamma$ Dor pulsators.

\subsection{Strongly magnetic SPB stars} \label{sec:mag_SPB}

The SPB stars are typically mid- to late-type B stars that pulsate in g modes excited by the $\kappa$ mechanism via the partial ionization of iron-group elements. The predicted range of radial orders for $l$ = 1 and 2 oscillations are between about $n$ = 10 and 80, which increases with age \citep{2002CoAst.142...10P, 2017MNRAS.469...13S}. Models for individual stars found typical values of $n \sim$ 5--30 \citep[e.g.,][]{2015ApJ...810...16T, 2019ApJ...881...86W, 2022MNRAS.511.1529S}. While the excitation mechanisms differ between SPB and  $\gamma$ Dor stars, their pulsations are similar in that they are intermediate- to high-radial-order g modes.

Several SPB stars have been confirmed to have a strong, organized magnetic field consistent with a fossil field, and there is currently no indication that the incidence rates of strong magnetism are any different for SPB stars compared to non-pulsating stars in the same mass regime. 
Examples include $\zeta$ Cas (= HD 3360), with a predominantly dipolar field and strength of $B_{\rm pol} = 335$ G \citep{2003A&A...406.1019N}; and 16 Peg (= HD 208057), with a similar magnetic field \citep{2009IAUS..259..393H, 2009MNRAS.398.1505S}. Another prominent example is the SPB\footnote{There are some higher frequency signals in the range of $\sim$4.5--6.5 d$^{-1}$ that were initially presumed to be p modes. However, these can be explained as rotationally shifted g modes, so HD 43317 is not considered a hybrid pulsator \citep{2018A&A...616A.148B}.} star HD 43317, with a dipolar field strength of approximately 1 kG \citep{2013A&A...557L..16B, 2017A&A...605A.104B, 2018A&A...616A.148B}. HD~43317 is an interesting case, as it has also been the subject of asteroseismic analysis to study both its interior structure and the strength of the magnetic field in the near-core region \citep{Lecoanet2022}. The main idea behind the work of \citet{Lecoanet2022} is that strong, near-core magnetism can act to suppress high-radial-order (which have lower frequency) g modes, and in HD 43317 only higher-frequency lower-radial order g modes are observed (see also Sect.~\ref{sec:internalMag}). Using an optimal stellar model from the asteroseismic analysis, it was determined that a near-core field with a strength of $\sim$5 $\times$ 10$^{5}$ G is compatible with suppressing the unobserved high-order g modes while remaining too weak to suppress the observed lower order g modes. In the case of HD 43317 it was determined that g-mode radial orders between $n$ = 1 and 15 are observed, with higher radial orders suppressed \citep{2018A&A...616A.148B, Lecoanet2022}.

Since there are several examples of strongly magnetic SPB stars despite the magnetic suppression of some modes, a plausible explanation for the lack of strongly magnetic $\gamma$ Dor pulsators is that a strong fossil field inhibits the excitation mechanism involving the outer convective layer (convective flux blocking). For SPB stars, magnetism does not seem to inhibit the excitation of g modes via the $\kappa$ mechanism.

\subsection{Possible reasons for our lack of magnetic detections in $\gamma$ Dor stars} \label{sec:possible_reasons}

There are several possible reasons that may explain the lack of strongly magnetic $\gamma$ Dor pulsators in our sample. These are listed and briefly described below.

\textit{\emph{Insufficient sample size and/or quality:}} The fraction of mCP stars (which are strongly magnetic as a class) is believed to be low in the mass range where $\gamma$ Dor pulsation is found \citep[][Sect.~\ref{sec:intro}]{2007pms..conf...89P, 2019MNRAS.483.2300S}. In a blind sample of 47 $\gamma$ Dor stars, it may thus be unlikely to find even one strongly magnetic mCP star. Our sample, however, was not blind---efforts were made to prioritize objects with signs of rotational modulation that should favor magnetic stars \citep{2019MNRAS.487..304D, 2023A&A...676A..55L}.
Additionally, our sample is biased toward slower rotators, which generally have higher rates of strong magnetic fields compared to more rapid rotators \citep{1995ApJS...99..135A}.
Indeed, of the three magnetic detections, two or three of them show rotational modulation (Sect.~\ref{sec:magnetic}; rotation is possible but unclear for 78 UMa) but lack $\gamma$ Dor pulsations in the magnetic component. However, given the practical limitations of spectropolarimetric observation (requiring bright and slowly rotating targets), it was very difficult to find $\gamma$ Dor stars with stable rotational modulation in the TESS photometry. 

\textit{\emph{Insufficient spectropolarimetric data quality:}} It is possible that, in at least a certain fraction of our sample, there are surface magnetic fields that are too weak for our detection limits. However, our upper limits (Sect.~\ref{sec:nonmagnetic}) are in almost all cases sufficient to rule out typical fossil fields as seen in other intermediate- and high-mass stars with typical strengths of 100 G and above (Sect.~\ref{sec:sample}). 

\textit{\emph{Strong global magnetism and $\gamma$ Dor pulsation are mutually exclusive:}} While typical fossil fields have strengths of 100 G and above at the surface, the field strength in the near core region is significantly higher \citep[e.g.,][]{2009ApJ...705.1000F, Lecoanet2022}. Strong near-core fields are capable of suppressing g-mode pulsation by converting the energy into Alfvén waves (Sects.~\ref{sec:internalMag} and~\ref{sec:mag_SPB}). Thus, g modes that are excited could be quickly damped and thus not observed. The existence of strongly magnetic SPB stars (Sect.~\ref{sec:mag_SPB}) demonstrates that strong fossil fields and intermediate- and high-radial-order g-mode pulsation can generally coexist; whether this is true or not for the lower mass $\gamma$ Dor stars remains to be tested. There is also the potential for interaction between a strong global field and the outer convection zone in $\gamma$ Dor stars \citep{2020ApJ...900..113J, 2019MNRAS.487.3904M, 2013MNRAS.433.2497S}. A strong magnetic field could interfere with the convective blocking mechanism that drives $\gamma$ Dor pulsation. The inverse could also be true; the thin convective zone may disrupt the outer field, confining the magnetic extent to more interior layers and thus being unobservable at the surface. This seems less likely, owing to the many strongly magnetic stars that occupy the same region of the HR diagram where $\gamma$ Dor stars are found. Testing these hypotheses would require detailed magneto-hydrodynamic simulations, which to our knowledge have not yet been carried out in this scenario.

\section{Conclusions} \label{sec:conclusions}

To date, there are no $\gamma$ Dor pulsators with a directly detected magnetic field at the stellar surface. 
Although our spectropolarimetric survey of 47 A- and F-type stars did result in three magnetic detections, none of these magnetic objects are consistent with hosting $\gamma$ Dor pulsations (Sects.~\ref{sec:mag_iotphe},~\ref{sec:mag_TYC}, and~\ref{sec:mag_78UMa}).
The remainder of the sample, entirely without detected surface fields, shows photometric variability generally consistent with $\gamma$ Dor pulsation, although in a small number of cases the variability may be better described by (variable) rotational modulation and/or low frequency signals in more evolved stars. 

Although we cannot provide a number stating the incidence rate of strong magnetism in $\gamma$ Dor pulsators in general, it is certainly low and consistent with zero. For the stars in our sample without a detected magnetic field (Sect.~\ref{sec:nonmagnetic}), we find upper limits for the dipolar field strength between 10 and 100 G at the 95\% credible regions (or between 5 and 40 G at the 68\% credible regions). There are only two exceptions, with 95\% credible region upper limits of 120 and 130 G (Sect.~\ref{sec:nonmagnetic}). 

Weak global magnetic fields on the order of a few Gauss or lower may still be present in a fraction of $\gamma$ Dors (Sect.~\ref{sec:weak_global_fields}). Likewise, variable dynamo-driven magnetism may emerge at the surface of $\gamma$ Dor stars with field strengths too weak to detect in our data. There is indirect evidence of dynamo-driven magnetism being present in some $\gamma$ Dor stars given that variable rotational modulation is occasionally observed in space photometry, especially from $Kepler$ (Sect.~\ref{sec:dynamo_fields}). Asteroseismic analysis has proven to be a viable method to infer internal magnetic properties of $\gamma$ Dor stars, as demonstrated in \citet{Takata2026}. The upper limits for dipolar surface magnetic-field strengths as determined in this work provide crucial constraints for asteroseismic modeling with regard to magnetism. Such models (for $\gamma$ Dor stars) should not have a surface dipolar field strength above 100 G, which is a conservative estimate based on the 95\% credible regions for our spectropolarimetric sample. An upper limit of $\sim$50 G should still be a safe assumption, as there are no stars in our sample with an upper limit greater than 40 G at the 68\% credible region, and the mean value of the 95\% credible region's upper limit is 53 G. 

The emerging hypothesis is that strong fossil fields suppress the mode excitation mechanism that would otherwise operate in $\gamma$ Dor stars. Given the existence of strongly magnetic, g-mode pulsating SPB stars (Sect.~\ref{sec:mag_SPB}), magnetic mode damping seems unlikely to fully explain the lack of strongly magnetic $\gamma$ Dor pulsators. Likewise, the large number of stars in and near the $\gamma$ Dor instability strip (Sect.~\ref{sec:discussion_strongly_magnetic}) demonstrates that the outer convective zones expected in this stellar regime do not in general disrupt the outer regions of strong global fields. Detailed magneto-hydrodynamic simulations are required to study the nature of how a strong global magnetic field might interact with the outer convective zone in such a way that inhibits the convective blocking mechanism that drives pulsations in $\gamma$ Dor stars.

\section{Data availability}
The reduced and continuum-normalized spectropolarimetric datasets, the line masks and the excluded regions used to compute the LSD profiles, and the details of the magnetic measurements are available at \url{https://zenodo.org/records/21627429}. All TESS data products used in this work are publicly available e.g. on the MAST database (\url{https://mast.stsci.edu/}). 

\begin{acknowledgements}
The authors thank the anonymous referee for their thorough review and their numerous helpful comments that improved the quality and clarity of this manuscript. 
Based on observations obtained at the Canada-France-Hawaii Telescope (CFHT) which is operated by the National Research Council of Canada, the Institut National des Sciences de l'Univers of the Centre National de la Recherche Scientifique of France, and the University of Hawaii. 
Funded/Co-funded by the European Union (ERC, MAGNIFY, Project 101126182 ). Views and opinions expressed are however those of the author(s) only and do not necessarily reflect those of the European Union or the European Research Council. Neither the European Union nor the granting authority can be held responsible for them. This work was supported by the ``Action Spécifique de Physique Stellaire'' (ATPS) of CNRS/INSU co-funded by CEA and CNES. This work was supported by the scientific council of the Paris Observatory through the ``Action Incitative de Physique Stellaire (AIPS)''. P.S. acknowledges the Delaware Space Grant College and Fellowship Program (NASA Grant 80NSSC20M0045). This work has made use of the VALD database, operated at Uppsala University, the Institute of Astronomy RAS in Moscow, and the University of Vienna. This paper includes data collected by the TESS mission, which are publicly available from the Mikulski Archive for Space Telescopes (MAST). Funding for the TESS mission is provided by NASA's Science Mission directorate. This work has made use of data from the European Space Agency (ESA) mission {\it Gaia} (\url{https://www.cosmos.esa.int/gaia}), processed by the {\it Gaia} Data Processing and Analysis Consortium (DPAC \url{https://www.cosmos.esa.int/web/gaia/dpac/consortium}). Funding for the DPAC has been provided by national institutions, in particular the institutions participating in the {\it Gaia} Multilateral Agreement. This research has made use of NASA's Astrophysics Data System. This research has made use of the SIMBAD database, operated at CDS, Strasbourg, France. This research made use of Lightkurve, a Python package for Kepler and TESS data analysis \citep{Lightkurve2018}. This research made use of Astropy\footnote{http://www.astropy.org}, a community-developed core Python package for Astronomy \citep{astropy2013, astropy2018}.
Based on observations obtained at the Canada-France-Hawaii Telescope (CFHT) which is operated by the National Research Council of Canada, the Institut National des Sciences de l'Univers of the Centre National de la Recherche Scientifique of France, and the University of Hawaii. CFHT is located on Maunakea on Hawaii Island, a mountain of considerable cultural, natural, and ecological significance. Maunakea is a sacred site to Native Hawaiians, also known as K\'anaka\'oiwi. We would like to thank the Canada-France-Hawaii Telescope (CFHT) Operations and Software Groups for their contributions and diligence in maintaining observatory operations; the CFHT Astronomy Group for their observation coordination and data acquisition efforts; and the CFHT Finance \& Administration Group for their contributions to the management and administration of the observatory.
\end{acknowledgements}

\bibliographystyle{bibtex/aa} 
\bibliography{bibfile}

\clearpage

\begin{appendix}
\onecolumn

\section{Data tables} \label{sec:tbls}

Table~\ref{tab:specpol} lists the targets and the spectropolarimetric observing dates, exposure times, and spectroscopic signal to noise. Table~\ref{tab:results} gives the stellar parameters from the SED fitting and spectroscopic fitting routines, lists the magnetic fields for the four targets with detections, and provides dipolar magnetic field upper limits for the remainder of the sample. 

{
\begin{centering}
\begin{table}[h!]
\caption{Summary of ESPaDOnS spectropolarimetric observations for the sample.}
\resizebox{0.8\textwidth}{!}{
\begin{tabular}{cccccccc}
\hline
ID         & TIC    & ST         & Date & HJD-2450000 & exposure (s)  & S/N  \\
\hline
13 LMi          & 4752246    & F3V         & Jan14 2025 & 10691.0087  & 4x 18   & 185  \\
32 Tau          & 14156987   & F2          & Oct20 2023 & 10239.0107  & 4x 30   & 321  \\
37 UMa          & 289715844  & F1V         & Jan14 2025 & 10691.0570  & 4x 30   & 399  \\
39 And          & 352409254  & kA3hA7VmA9  & Aug26 2024 & 10550.0435  & 4x 165  & 656  \\
43 Cyg          & 277234493  & F0V         & Nov15 2019 & 8803.7343   & 4x 39   & 345  \\
               &            &             & May26 2023 & 10092.1315  & 4x 32   & 314  \\
47 Cet          & 423700466  & F0.5V       & Aug26 2024 & 10550.0537  & 4x 26   & 313  \\
51 Eri          & 298810162  & F0IV        & Aug26 2024 & 10550.1054  & 2x 4x 145  & 1027  \\
59 Dra          & 235682463  & A9V         & May29 2023 & 10095.0680  & 4x 111  & 220  \\
               &            &             & Oct24 2023 & 10242.6880  & 4x 150  & 695  \\
67 UMa          & 141201605  & F0          & Feb09 2023 & 9986.0659   & 4x 271  & 1022 \\
78 UMa          & 229534764  & F2V         & Jan14 2025 & 10691.0843  & 2x 4x 170  & 1496 \\
7 And           & 252756054  & F1V         & Aug26 2024 & 10550.0309  & 4x 104  & 962  \\
9 Aur           & 327701133  & F2V         & Aug26 2024 & 10550.0855  & 4x 14   & 258  \\
BD+07 442       & 387515681  & A5          & Sep12 2019 & 8740.1223   & 4x 1554 & 349  \\
$\delta$ Aql         & 228896480  & F1IV-V      & Aug26 2024 & 10549.9175  & 2x 4x 42   & 1491 \\
$\eta$ Ant         & 45883663   & F1V         & Jan14 2025 & 10691.0433  & 4x 102  & 660  \\
FO Cet          & 423431660  & F0V         & Sep13 2019 & 8741.1238   & 4x 780  & 759  \\
HD 102590       & 14724795   & A8          & Jan18 2024 & 10329.1149  & 4x 483  & 1085 \\
HD 109799       & 60709182   & F2IV        & Jan14 2025 & 10691.0961  & 4x 40   & 400  \\
HD 111456       & 142277151  & F6V         & Jan14 2025 & 10691.0718  & 4x 67   & 409  \\
HD 127821       & 166178883  & F5          & Aug22 2024 & 10545.7823  & 4x 279  & 487  \\
HD 145872       & 162519062  & F5          & Aug22 2024 & 10545.8361  & 4x 278  & 204  \\
HD 173109       & 351800186  & F0          & Apr08 2025 & 10774.9696  & 4x 143  & 174  \\
HD 184875       & 270610122  & A2V         & Nov15 2019 & 8803.6984   & 3x 4x 292  & 1993 \\
HD 196195       & 276400050  & F2          & Aug10 2025 & 10898.8186  & 4x 1762 & 1045 \\
HD 218396       & 245368902  & F0+VkA5mA5  & Aug26 2024 & 10550.0256  & 4x 65   & 397  \\
HD 2421         & 190996748  & A2Vs+F0     & Aug26 2024 & 10550.0369  & 4x 21   & 334  \\
HD 37594        & 11295159   & A9V         & Aug26 2024 & 10550.1145  & 4x 25   & 211  \\
HD 62094        & 141759686  & F6V         & Jan14 2025 & 10690.9976  & 4x 234  & 229  \\
HD 88815        & 142784946  & F2V         & Jan15 2024 & 10326.0563  & 4x 141  & 391  \\
               &            &             & Jan14 2025 & 10691.0316  & 4x 212  & 554  \\
HD 92787        & 150268783  & F5III       & Jan14 2025 & 10691.0647  & 4x 192  & 931  \\
HD 9335         & 136843852  & A5/7III     & Sep16 2019 & 8744.0004   & 4x 1331 & 709  \\
i Cen           & 30068122   & F2V         & Jan14 2025 & 10691.1004  & 4x 83   & 957  \\
$\iota$ Phe         & 144276313  & A2/ApSrCrEu & Oct25 2023 & 10243.8406  & 4x 20   & 360  \\
               &            &             & Dec30 2023 & 10309.6831  & 4x 20   & 361  \\
I pup           & 134497068  & F3          & Dec30 2023 & 10309.9798  & 4x 70   & 643  \\
k Ori           & 437886584  & F5V         & Aug26 2024 & 10550.1170  & 4x 14   & 259  \\
$\lambda$ CrB         & 29385285   & F2IV-V      & Aug22 2024 & 10545.8101  & 2x 4x 186  & 949  \\
$\mu$ Vir          & 185284200  & F2V         & Aug22 2024 & 10545.7601  & 4x 27   & 345  \\
$\nu$ Her          & 22127552   & F2II        & Jan20 2025 & 10697.1777  & 4x 12   & 345  \\
$\nu$ Per          & 431927240  & F4II        & Aug26 2024 & 10550.0834  & 4x 17   & 519  \\
NY UMa          & 18066482   & Am          & Nov20 2019 & 8809.1661   & 4x 96   & 183  \\
PR Peg          & 265720508  & A8V         & Sep15 2019 & 8742.9924   & 4x 259  & 143  \\
               &            &             & Sep16 2019 & 8743.9520   & 2x 4x 259  & 140   \\
TYC2 430-1205-1 & 172414656  & A6VSrEu     & Dec29 2023 & 10308.8307  & 4x 655  & 140  \\
V1006 Cas       & 201697061  & F0          & Nov16 2019 & 8804.8908   & 4x 465  & 138  \\
               &            &             & Nov18 2019 & 8806.8811   & 2x 4x 465  & 311  \\
V1012 Cas       & 240931244  & F2V         & Sep15 2019 & 8743.0623   & 2x 4x 1300 & 892  \\
V350 CMa        & 48881783   & F2V         & Jan14 2025 & 10690.9836  & 4x 181  & 468  \\
V372 Peg        & 384957100  & F3V         & Nov16 2019 & 8804.8041   & 2x 4x 10   & 82   \\
               &            &             & Nov18 2019 & 8806.7868   & 4x 10   & 84   \\
               &            &             & Aug26 2024 & 10550.0176  & 4x 14   & 167  \\
V418 Peg        & 60837939   & A3m         & Sep15 2019 & 8742.9774   & 2x 4x 79   & 124   \\
               &            &             & Nov16 2019 & 8804.8150   & 2x 4x 79   & 147  \\
$\zeta$ Leo         & 95360236   & F0IIIa      & Jan14 2025 & 10691.0222  & 2x 4x 37   & 1344  \\
\hline
\end{tabular}}
\label{tab:specpol}
\end{table}
\end{centering}
}

{\onecolumn
\begin{centering}
\begin{table}[ht!]
\caption{Summary of stellar parameters and magnetic measurements.}
\resizebox{0.98\textwidth}{!}{%
\begin{tabular}{ccccccccccccc}
\hline
ID             & ST         & T$_{\rm eff, SED}$ & L$_{\rm SED}$ & R$_{\rm SED}$  & T$_{\rm eff, spec}$  & log $g_{\rm spec}$  & $v$sin$i_{\rm spec}$ & Bin  &  B$_{max, dipole}$ &  B$_{max, dipole}$ & B$_{l}$    &  FAP\\
               &            &  (K)               & (L$_{\odot}$) & (R$_{\odot}$)  &       (K)             & (c.g.s)            & (km s$^{-1}$)   &    &  (G; 95\%) &   (G; 68\%) & (G)  &   \\
\hline
78 UMa          & F2V         & 6713   & 5.5    & 1.8   & 6692$\pm$254   & 3.96$\pm$0.35  & 83.0$\pm$4.5   & SB2  &        &        & 7$\pm$1& 1.7e-07 \\
BD+07 442       & A5          & 7398   & 8.3    & 1.8   & 6701$\pm$250   & 3.75$\pm$0.31  & 36.7$\pm$14.6  & ...  &        &        & 56$\pm$24& 0.0e+00\\
$\iota$ Phe     & A2/ApSrCrEu & 8335   & 61.5   & 4.1   & 7824$\pm$525   & 3.30$\pm$0.64  & 31.6$\pm$6.8   &  ... &        &        & 71$\pm$12& 4.3e-09\\
     &   &     &     &     &     &    &     &    &        &        & 45$\pm$13& 0.0e+00\\
TYC 2430-1205-1 & A6VSrEu     & 8297   & 28.5   & 2.7   & 6705$\pm$327   & 2.90$\pm$0.38  & 54.7$\pm$16.9  &  SB2 &        &        & 132$\pm$44& 9.5e-10\\
\hline
13 LMi          & F3V         & 6826   & 8.2    & 2.1   & 6521$\pm$178   & 3.73$\pm$0.23  & 21.1$\pm$0.8   &  ... & $<$73  & $<$18  & 7$\pm$8& 8.0e-02\\
32 Tau          & F2          & 6936   & 8.3    & 2.0   & 6672$\pm$202   & 3.78$\pm$0.25  & 20.3$\pm$1.1   &  ... & $<$27  & $<$8   & 1$\pm$4& 4.5e-01\\
37 UMa          & F1V         & 7000   & 4.8    & 1.5   & 6800$\pm$193   & 4.03$\pm$0.24  & 42.9$\pm$2.9   &  ... & $<$91  & $<$25  & -10$\pm$12& 6.3e-02\\
39 And          & kA3hA7VmA9  & 8006   & 37.1   & 3.3   & 7028$\pm$411   & 3.22$\pm$0.64  & 42.0$\pm$5.4   & ...  & $<$26  & $<$7   & 1$\pm$7& 2.3e-01\\
43 Cyg          & F0V         & 7015   & 5.8    & 1.7   & 6606$\pm$196   & 3.73$\pm$0.27  & 42.7$\pm$2.1   &  ... & $<$41  & $<$12  & 4$\pm$12& 9.3e-01\\
          &           &     &     &    &   &    &     &    &   &    & -13$\pm$14& 5.4e-02\\
47 Cet          & F0.5V       & 7178   & 7.0    & 1.7   & 6984$\pm$213   & 4.08$\pm$0.25  & 20.3$\pm$1.6   &  ... & $<$28  & $<$7   & -1$\pm$5& 2.7e-01\\
51 Eri          & F0IV        & 7441   & 6.0    & 1.5   & 7009$\pm$262   & 3.98$\pm$0.35  & 63.2$\pm$4.5   &  ... & $<$55  & $<$15  & 2$\pm$8& 5.6e-01\\
59 Dra          & A9V         & 6998   & 5.4    & 1.6   & 6990$\pm$244   & 4.10$\pm$0.31  & 52.2$\pm$3.0   & ...  & $<$44  & $<$11  & 7$\pm$10& 7.2e-02\\
          &           &     &     &     &     &    &     &    &    &    & -9$\pm$31& 6.1e-01\\
67 UMa          & F0          & 7529   & 7.7    & 1.7   & 6916$\pm$313   & 3.63$\pm$0.46  & 69.2$\pm$6.1   & ...  & $<$59  & $<$17  & -5$\pm$9& 1.3e-01\\
7 And           & F1V         & 7112   & 7.5    & 1.8   & 7038$\pm$298   & 4.06$\pm$0.39  & 57.6$\pm$3.9   & ...  & $<$26  & $<$7   & 8$\pm$8& 9.4e-01\\
9 Aur           & F2V         & 6931   & 5.5    & 1.7   & 6920$\pm$188   & 4.07$\pm$0.22  & 19.9$\pm$1.5   & ...  & $<$42  & $<$11  & 5$\pm$6& 8.5e-01\\
$\delta$ Aql    & F1IV-V      & 7005   & 8.9    & 2.0   & 6780$\pm$262   & 3.50$\pm$0.41  & 81.4$\pm$3.6   &  SB? &  $<$28  &  $<$8 & -3$\pm$8& 3.7e-01\\
$\eta$ Ant      & F1V         & 7254   & 7.9    & 1.8   & 6935$\pm$275   & 3.93$\pm$0.35  & 46.7$\pm$3.0   &  ... & $<$35  & $<$9   & -1$\pm$8& 7.8e-01\\
FO Cet          & F0V         & 7138   & 7.2    & 1.8   & 7533$\pm$198   & 4.52$\pm$0.23  & 59.4$\pm$6.0   &  ... & $<$121 & $<$32  & -14$\pm$14& 4.0e-01\\
HD 102590       & A8          & 7131   & 12.8   & 2.4   & 6892$\pm$278   & 3.80$\pm$0.38  & 76.1$\pm$2.8   &  SB3 &  $<$39   &   $<$10  & 1$\pm$10& 5.3e-03\\
HD 109799       & F2IV        & 6978   & 6.3    & 1.7   & 6692$\pm$158   & 3.73$\pm$0.20  & 39.0$\pm$1.8   &  ... & $<$50  & $<$14  & -8$\pm$10& 7.6e-01\\
HD 111456       & F6V         & 6327   & 2.2    & 1.2   & 6382$\pm$165   & 4.11$\pm$0.21  & 41.1$\pm$2.7   &  ... &  $<$43 & $<$11  & -12$\pm$10& 2.0e-03\\
HD 127821       & F5          & 6604   & 3.0    & 1.3   & 6623$\pm$150   & 4.16$\pm$0.19  & 51.9$\pm$4.0   &  ... & $<$41  & $<$11  & 15$\pm$13& 9.8e-01\\
HD 145872       & F5          & 6102   & 2.7    & 1.5   & 6186$\pm$128   & 3.89$\pm$0.17  & 28.6$\pm$1.2   &  ... & $<$55  & $<$15  & 4$\pm$10& 8.1e-01\\
HD 173109       & F0          & 7003   & 12.5   & 2.6   & 7074$\pm$232   & 4.08$\pm$0.26  & 13.0$\pm$1.0   &  ... & $<$22  & $<$6   & 0$\pm$5& 6.4e-01\\
HD 184875       & A2V         & 8433   & 191.6  & 8.1   & 8867$\pm$202   & 4.80$\pm$0.21  & 89.2$\pm$8.0   &  SB? & $<$130 & $<$38  & -24$\pm$16& 6.4e-01\\
HD 196195       & F2          & 6900   & 12.9   & 2.8   & 6414$\pm$277   & 3.14$\pm$0.46  & 87.0$\pm$5.8   &  ... & $<$36  & $<$10  & -4$\pm$9& 8.5e-01\\
HD 218396       & F0+VkA5mA5  & 7322   & 6.2    & 1.6   & 7977$\pm$142   & 4.90$\pm$0.11  & 35.5$\pm$2.7   &  ... & $<$55  & $<$14  & 5$\pm$14& 3.6e-01\\
HD 2421         & A2Vs+F0     & 8917   & 60.8   & 3.4   & 9494$\pm$238   & 3.65$\pm$0.34  & 8.0$\pm$9.3    &  SB2 &  $<$27  &  $<$6  & -2$\pm$3& 6.8e-01\\
HD 37594        & A9V         & 7222   & 5.9    & 1.6   & 7255$\pm$222   & 4.36$\pm$0.25  & 17.8$\pm$2.0   &  ... & $<$59  & $<$18  & 7$\pm$7& 6.3e-01\\
HD 62094        & F6V         & 6612   & 3.3    & 1.5   & 6492$\pm$119   & 4.00$\pm$0.15  & 25.3$\pm$0.9   &  ... & $<$57  & $<$15  & 4$\pm$9& 5.9e-01\\
HD 88815        & F2V         & 6955   & 8.7    & 2.1   & 6975$\pm$236   & 3.98$\pm$0.29  & 41.6$\pm$2.5   &  SB2 &  $<$20   &   $<$6 & 10$\pm$9& 2.0e-01\\
         &           &     &      &     &     &    &     &    &      &     & 10$\pm$12& 9.5e-01\\
HD 92787        & F5III       & 7129   & 9.2    & 2.0   & 7042$\pm$240   & 3.94$\pm$0.31  & 55.4$\pm$3.5   &  ... & $<$33  & $<$9   & -3$\pm$8& 8.6e-01\\
HD 9335         & A5/7III     & 7734   & 13.3   & 2.1   & 6708$\pm$270   & 3.13$\pm$0.46  & 55.6$\pm$3.5   &  ... & $<$64  & $<$21  & -5$\pm$10& 8.3e-01\\
i Cen           & F2V         & 6800   & 5.9    & 1.8   & 6531$\pm$176   & 3.78$\pm$0.24  & 59.8$\pm$3.5   &  ... & $<$29  & $<$8   & -8$\pm$7& 1.3e-01\\
I pup           & F3          & 6854   & 6.1    & 1.8   & 7261$\pm$231   & 4.37$\pm$0.28  & 49.6$\pm$3.9   &  ... & $<$42  & $<$11  & -3$\pm$11& 5.2e-01\\
k Ori           & F5V         & 6641   & 2.9    & 1.3   & 6450$\pm$108   & 4.01$\pm$0.14  & 20.7$\pm$0.8   &  ... & $<$19  & $<$5   & 6$\pm$6& 5.1e-01\\
$\lambda$ CrB   & F2IV-V      & 7005   & 9.7    & 2.2   & 6794$\pm$217   & 3.71$\pm$0.31  & 73.2$\pm$2.8   &  ... & $<$49  & $<$14  & 4$\pm$10& 4.7e-01\\
$\mu$ Vir       & F2V         & 6672   & 7.6    & 2.1   & 6743$\pm$167   & 3.95$\pm$0.21  & 44.9$\pm$2.9   &  ... & $<$90  & $<$29  & 17$\pm$16& 4.2e-01\\
$\nu$ Her       & F2II        & 6571   & 1056.3 & 33.0  & 6787$\pm$281   & 3.61$\pm$0.39  & 24.8$\pm$1.4   &  ... & $<$21  & $<$5   & 1$\pm$5& 8.6e-01\\
$\nu$ Per       & F4II        & 6416   & 834.1  & 29.5  & 6171$\pm$183   & 2.74$\pm$0.28  & 44.3$\pm$2.3   &  SB? & $<$87  & $<$5   & 2$\pm$6& 2.5e-02\\
NY UMa          & Am          & 7157   & 6.2    & 1.7   & 6908$\pm$233   & 4.02$\pm$0.27  & 22.0$\pm$1.3   &  SB2 &  $<$46  &  $<$12  & -4$\pm$9& 2.9e-01\\
PR Peg          & A8V         & 6728   & 5.6    & 1.8   & 6728$\pm$194   & 3.90$\pm$0.23  & 18.1$\pm$1.5   &  SB? & $<$64  & $<$19  & 8$\pm$11& 5.7e-01\\
           &           &     &     &     &    &    &     &    &    &    & 7$\pm$9& 2.0e-01\\
V1006 Cas       & F0          & 7172   & 5.1    & 1.5   & 6532$\pm$210   & 3.58$\pm$0.30  & 41.1$\pm$1.9   &  SB2 &  $<$78 &  $<$22 & -1$\pm$14& 7.9e-01\\
       &           &     &      &     &     &    &    &    &    &    & 11$\pm$34& 1.8e-01\\
V1012 Cas       & F2V         & 6800   & 8.6    & 2.2   & 6863$\pm$212   & 3.96$\pm$0.28  & 88.7$\pm$5.3   &  SB2 &  $<$92  &  $<$24  & -20$\pm$24& 5.4e-01\\
V350 CMa        & F2V         & 7161   & 7.6    & 1.8   & 6881$\pm$221   & 3.92$\pm$0.28  & 49.0$\pm$2.6   &  ... & $<$65  & $<$17  & -4$\pm$12& 3.3e-02\\
V372 Peg        & F3V         & 7015   & 6.5    & 1.8   & 6614$\pm$162   & 3.80$\pm$0.19  & 9.5$\pm$0.4    &  SB? & $<$15  & $<$4   & -9$\pm$6& 4.3e-02\\
         &           &     &     &    &     &    &      &    &    &     & -1$\pm$6& 3.6e-01\\
         &           &     &     &    &     &    &      &    &    &     & -1$\pm$2& 8.2e-01\\
V418 Peg        & A3m         & 6397   & 11.0   & 2.9   & 6758$\pm$276   & 3.42$\pm$0.39  & 23.7$\pm$5.0   &  SB2 &   $<$101 &  $<$28 & -1$\pm$23& 1.1e-02\\
         &           &     &     &     &     &    &     &    &     &    & 9$\pm$21& 5.8e-02\\
$\zeta$ Leo     & F0IIIa      & 6720   & 231.5  & 12.0  & 6612$\pm$267   & 3.03$\pm$0.45  & 71.8$\pm$4.6   &  SB? & $<$95  & $<$5   & -9$\pm$6& 4.4e-05\\
\hline
\end{tabular}}
\label{tab:results}
\tablefoot{ The top four entries are for systems with detected magnetism, and the remainder include magnetic dipole upper limits at the 95\% and 68\% credible regions. Systems that are spectroscopic binaries (SB) are indicated, including candidates (``SB?''). For binaries, the magnetic upper limits listed are for the primary component. For 78 UMa, the magnetic detection corresponds to the secondary. Uncertainties on the SED-derived quantities are approximately 5\% for T$_{\rm eff, SED}$ 8\% for R$_{\rm SED}$, and 35\% for L$_{\rm SED}$ (see Sect.~\ref{sec:stellar_parameters}). Spectroscopic quantities that may not be accurate due to poor matches between synthetic spectra and the observed data are marked with an asterisk -- these are primarily for chemically peculiar stars. In some cases, $\log g$ seems unrealistically high, despite good agreement between synthetic and observed spectra. For BD+07 2442, the listed SED quantities were not determined by us due to a lack of parallax information, but adopted from \citet{2017AstBu..72...51S}. BD+07 2442 does not show pulsation but is included here because it was part of our spectropolarimetric survey and a strong field was detected. }
\end{table}
\end{centering}
}

\section{Stars with additional pulsation signals} \label{sec:other_pulsation}

\begin{figure}
    \centering
    \includegraphics[width=0.5\textwidth]{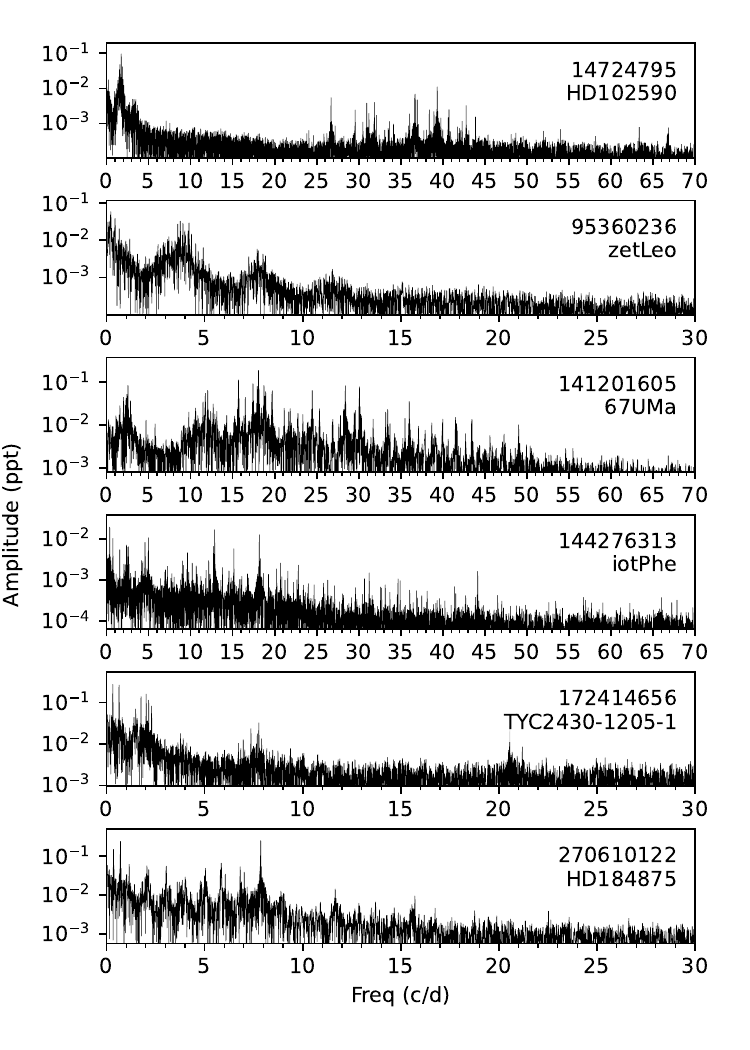}
    \caption{TESS frequency spectra of objects with high-frequency signals.}
    \label{fig:dSct}
\end{figure}

Frequency spectra of six stars with additional higher-frequency signals are shown in Fig.~\ref{fig:dSct}. Four of these (HD 102590, 67 UMa, $\iota$ Phe, and TYC 2430-1205-1) seem consistent with $\delta$ Scuti pulsation. The other two objects, HD 184875 and $\zeta$ Leo, display atypical frequency patterns. These objects are briefly discussed in the following sub-sections, except for $\iota$ Phe and TYC 2430-1205-1 which are strongly magnetic and discussed in the main text (Sections~\ref{sec:mag_iotphe},~\ref{sec:mag_TYC}).

\subsection{HD 184875 (=TIC 270610122)} \label{sec:hd184875}
HD~184875 exhibits clear rotational modulation in its TESS photometry (f$_{\rm rot}$ = 0.3548 d$^{-1}$) as well as low-frequency signals consistent with g-mode pulsation, and was thus a good candidate for searching for surface magnetism. The non-rotational signals are somewhat unusual, as there are approximately seven groups of frequencies, each spaced by about 1 d$^{-1}$, between about 1 and 8 d$^{-1}$. This does not follow the typical patterns exhibited by $\gamma$ Dor or $\delta$ Scuti pulsators, but these signals almost certainly represent pulsations of some type. The spectral type (A2V) is consistent with the parameters derived from the SED and spectral line fitting. However, inspection of the Stokes I profile (Fig.~\ref{fig:LSDFT_test4}) suggests it may be a binary, with two somewhat rapidly rotating stars of similar luminosity. It is also possible that the line profile distortions are caused by pulsation. Even if the luminosity is over-estimated by a factor of $\sim$2, the star(s) may still be somewhat evolved considering the HRD position. If HD~184875 is a system with two stars of similar brightness, then the dipolar magnetic upper limits are under-estimated by a factor of $\sim$2.

\begin{figure*}
    \centering
    \includegraphics[width=0.95\textwidth]{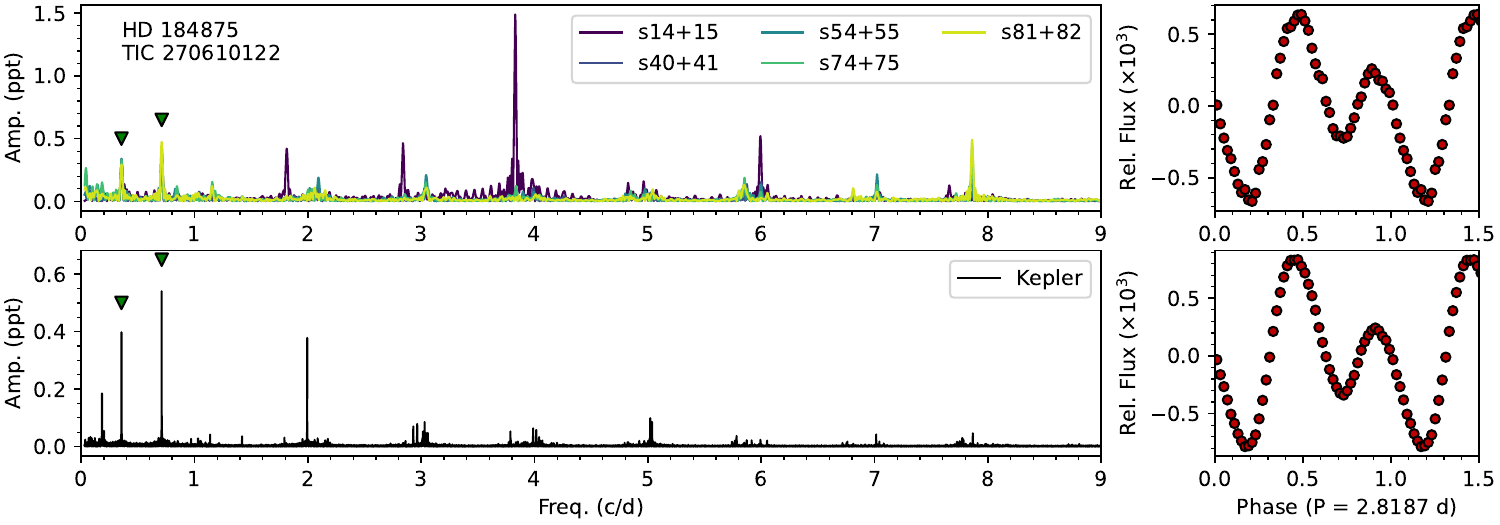}
    \caption{Data from TESS (top) and Kepler (bottom) for HD 184875. The different pairs of consecutive TESS sectors are plotted as different colors. The rotational frequency and its first harmonic are marked by triangles in the frequency spectra, and the light curves from the two satellites are plotted phased to the rotation period in the right panels, with 50 bins in phase.  }
    \label{fig:HD184875}
\end{figure*}

The rotational signal is constant over the TESS observing baseline. However, in the first two sectors where HD~184875 was observed by TESS (sectors 14+15), the non-rotational signals differ significantly from the later sectors, being in some places significantly higher in amplitude and at some frequencies with enhanced power not seen in other sectors. The later sectors (40, 41, 54, 55, 74, 75, 81, and 82) are all consistent with each other. This star was also observed by Kepler in the prime mission, with the light curve made available through the smear campaign \citep{2019ApJS..244...18P}. \citet{2019ApJS..244...18P} pointed out that HD 184875 "is a $\gamma$ Dor but also shows evidence for an unknown contaminant." It is not clear if all of the photometric signals seen in both Kepler and TESS originate in one star or in two, or what the nature of these signals are (except for the rotational signal which is unambiguous). There is no evidence for contamination from nearby stars on the sky. Fig.~\ref{fig:HD184875} compares the TESS and Kepler data, emphasizing the changing signals seen during TESS observations, as well as the stability of the rotational modulation signal in both datasets. The non-rotational variations are similar in both datasets, but with some clear differences.

\subsection{HD 102590 (=TIC 14724795)}
The TESS data for this star show both typical $\gamma$ Dor pulsation and $\delta$ Scuti pulsation (Fig.~\ref{fig:dSct}), and the stellar parameters we determined place it near the overlap of these two instability strips. It is known to be a close visual binary, with about 1 arcsecond between the components. The primary star (HD 102590A) is almost certainly the pulsator. The secondary (HD 102590B) is classified as G2. Inspection of the Stokes I profile shows a wide-lined primary and a ``double'' narrow-lined component, suggesting that the fainter secondary is likely a binary itself. 

\subsection{67 UMa (=TIC 141201605)}
The TESS photometry of 67 UMa shows low-frequency signals consistent with $\gamma$ Dor pulsation, as well as a rich set of higher-frequency $\delta$ Scuti signals (Fig.~\ref{fig:dSct}). While the Stokes I profile in our single observation has some structure, this may be pulsational in nature and does not necessarily indicate binarity. The star is commonly designated as Am in the literature \citep{2014yCat....1.2023S}. 67 UMa appears to be a hybrid $\gamma$ Dor and $\delta$ Scuti pulsator, but we cannot rule out the possibility of a pulsating companion star.

\subsection{$\zeta$ Leo (=TIC 95360236)} \label{sec:zetLeo}
This system presents the most complex Stokes I profile of the entire sample, potentially being made up of four or five components (Fig.~\ref{fig:LSDFT_test6}). The photometric variations include many low-frequency signals (below $\sim$1 d$^{-1}$), a wide ``hump'' made up of many closely-spaced signals (between about 2.5 -- 5 d$^{-1}$), and a second lower-amplitude hump at about twice the frequency of the first (Fig.~\ref{fig:dSct}). The somewhat unusual pulsational signals may play some part in distorting the Stokes I profile. High order multiplicity may also contribute to the Stokes I complexity. With only one spectropolarimetric snapshot, the nature of this system is unclear. $\zeta$ Leo appears as an evolved star according to the SED fitting, but this may be inaccurate if, e.g., there are several stars contributing to the SED. For the same reasons, our spectroscopic parameters are also questionable. The magnetic analysis of our single spectropolarimetric observation technically results in a marginal detection (MD) with a FAP of 4.4$\times$10$^{-5}$, and B$_{l}$ = -9$\pm$6 treating $\zeta$ Leo as a single star. Further spectroscopic observations are warranted to determine how or if the line profile varies with time to distinguish between multiplicity, rotation, and/or pulsation-induced line profile variations. Following this, deeper spectropolarimetric observations could better test for the presence of a magnetic field. With the data at hand, we do not claim a magnetic detection.

\section{LSD plots and TESS frequency spectra} \label{apx:LSD}

\begin{figure*}
    \centering
    \includegraphics[width=0.33\textwidth]{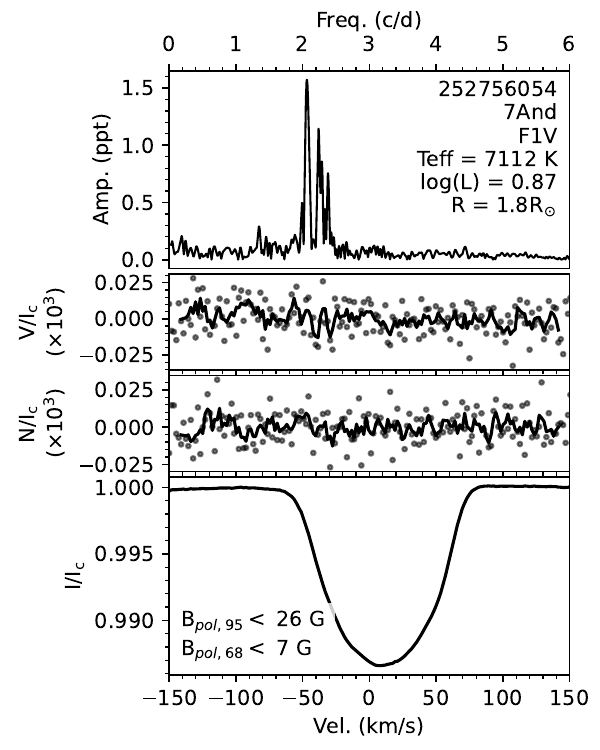}
    \includegraphics[width=0.33\textwidth]{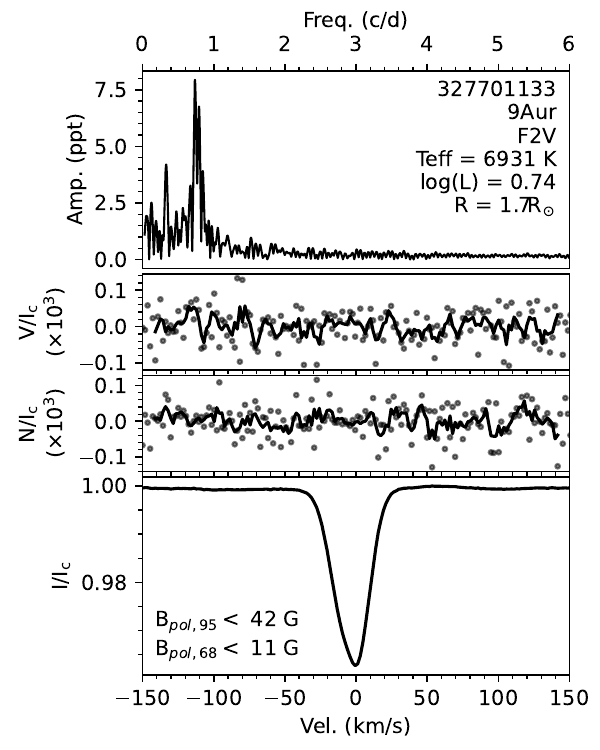}
    \includegraphics[width=0.33\textwidth]{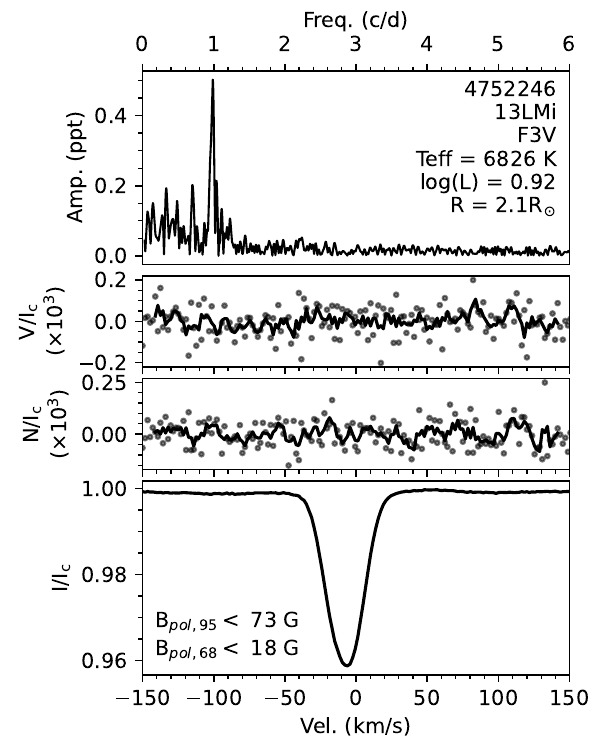}
    \includegraphics[width=0.33\textwidth]{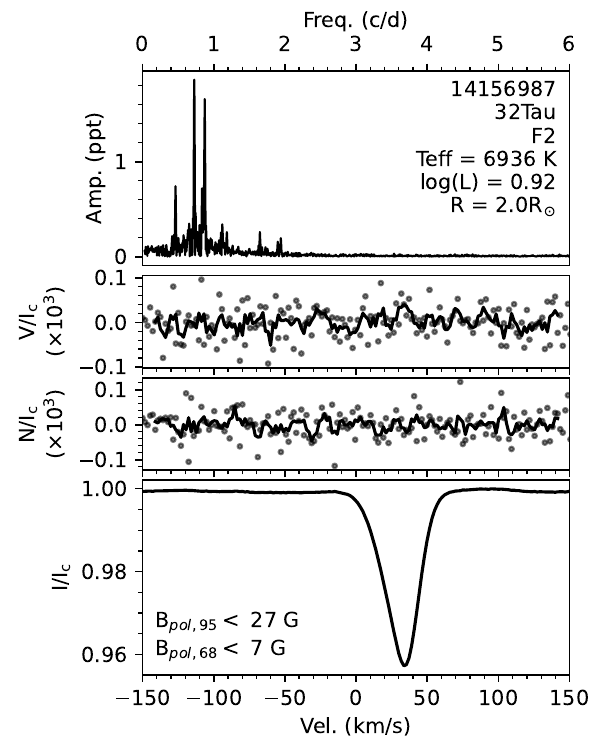}
    \includegraphics[width=0.33\textwidth]{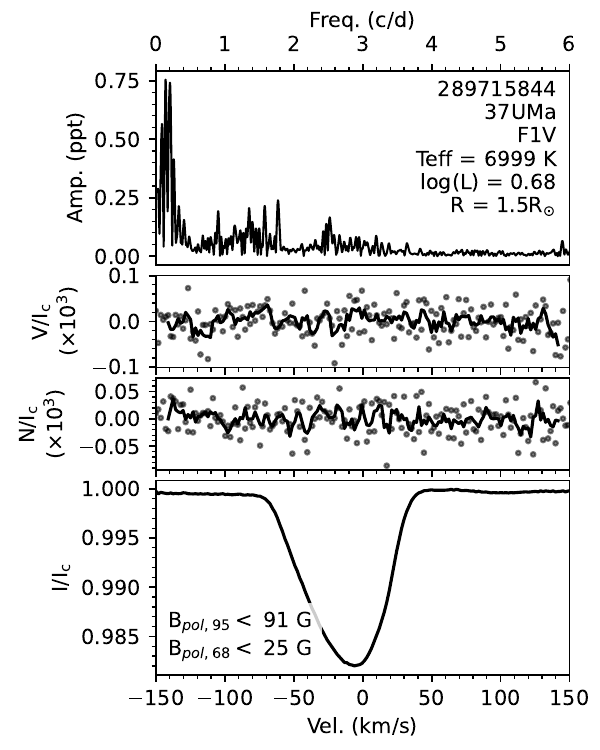}
    \includegraphics[width=0.33\textwidth]{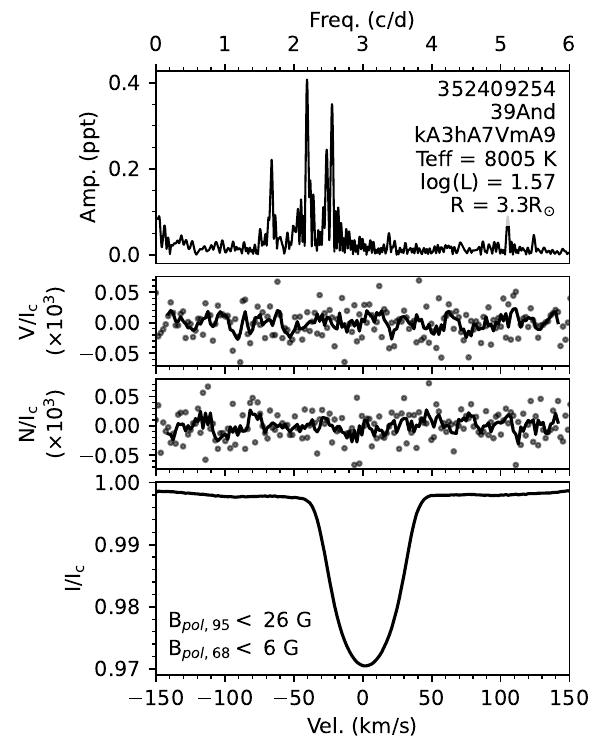}
    \includegraphics[width=0.33\textwidth]{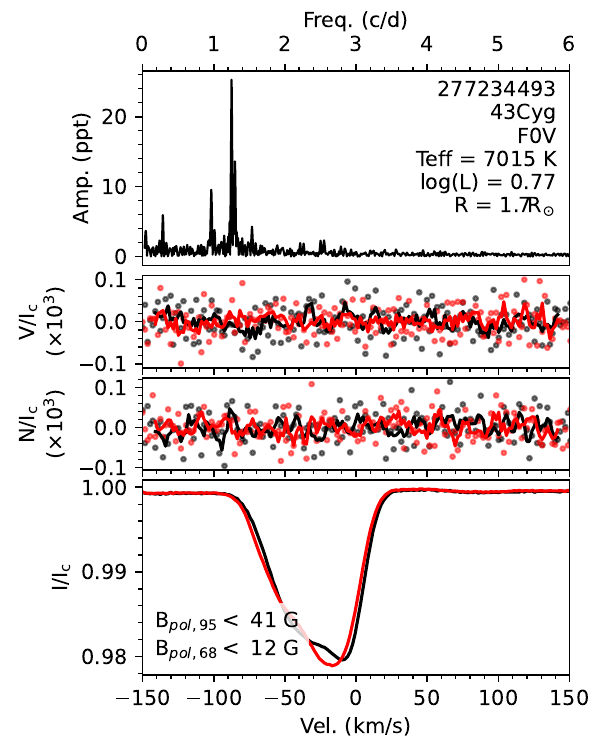}
    \includegraphics[width=0.33\textwidth]{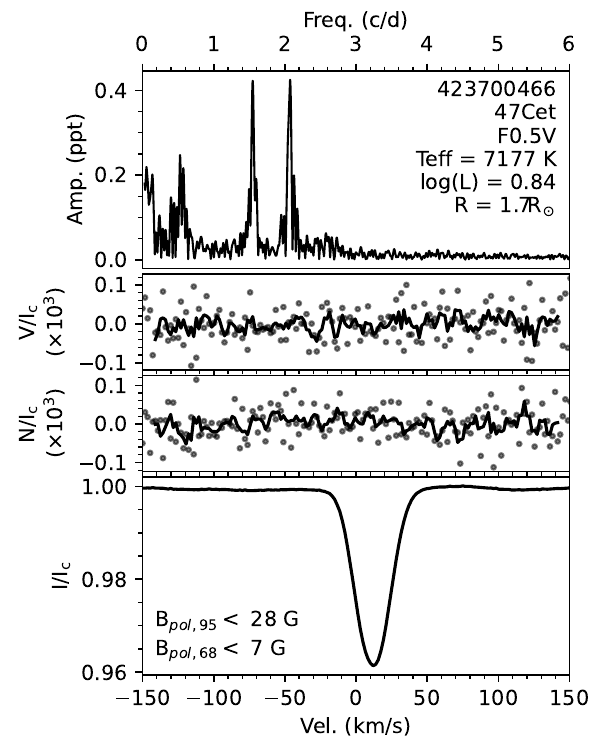}
    \includegraphics[width=0.33\textwidth]{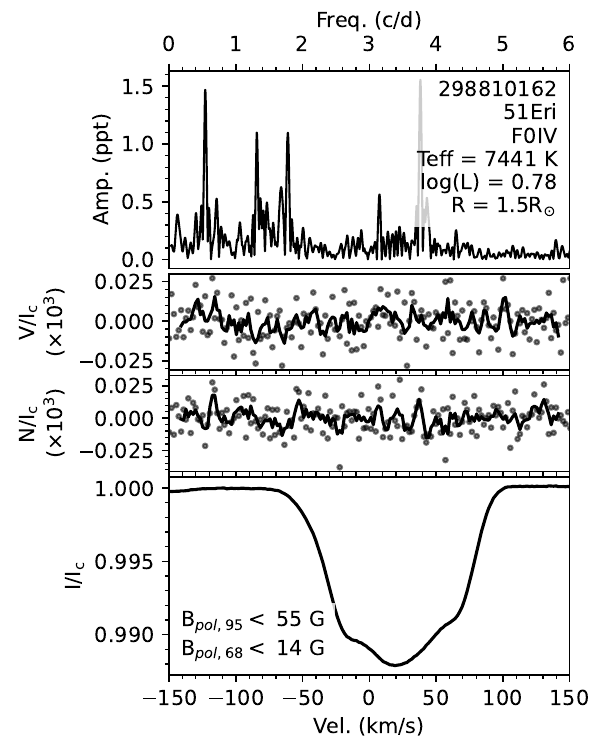}
    \caption{For each star, from top to bottom is shown the TESS frequency spectrum, Stokes $V$, Stokes $N$, and Stokes $I$. Whenever there are multiple spectropolarimetric observations, these are plotted in different colors. For the non-detections, upper limits at the 95\% and 68\% credible regions are shown. For obvious binaries, these values are given for both components. }
    \label{fig:LSDFT_test1}
\end{figure*}

The LSD profiles (Stokes $V$, $N$, and $I$) are plotted for all observations for the full sample in Fig.~\ref{fig:LSDFT_test1}. Included in these plots are a portion of the TESS frequency spectrum showing the low-frequency signals, and the stellar parameters T$_{\rm eff}$, log($L$), and radius (R) determined from the SED fitting, as well as the detected magnetic field strength or the dipolar magnetic field upper limits in the case of non-detections. Different colors in the Stokes profiles are used whenever there are two or three spectropolarimetric datasets.

\renewcommand{\thefigure}{C.\arabic{figure} (Cont.)}
\addtocounter{figure}{-1}

\begin{figure*}
    \centering
    \includegraphics[width=0.33\textwidth]{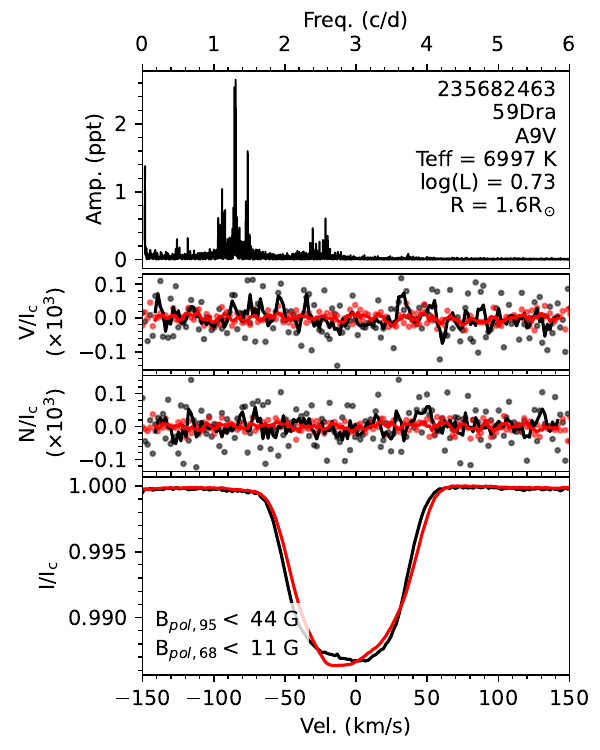}
    \includegraphics[width=0.33\textwidth]{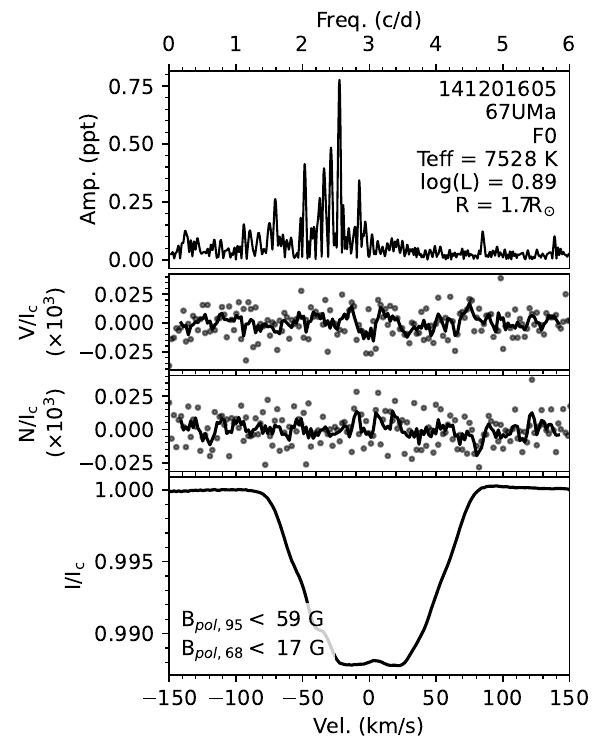}
    \includegraphics[width=0.33\textwidth]{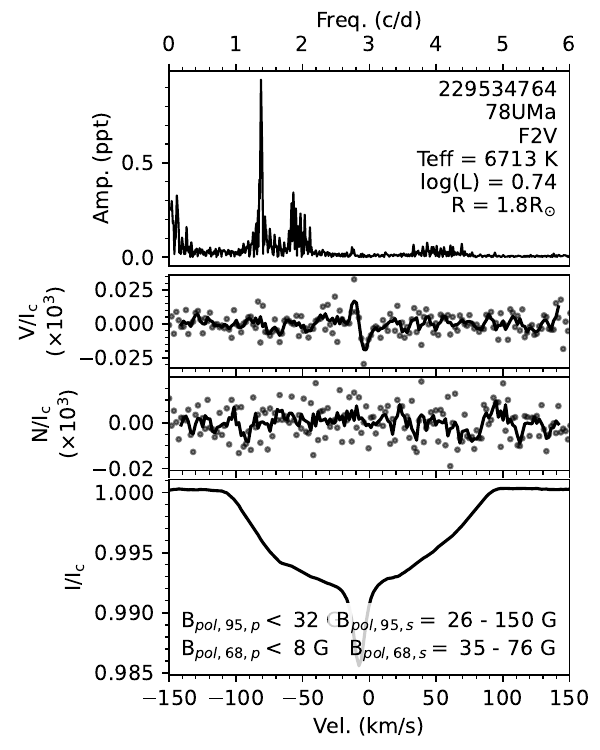}
    \includegraphics[width=0.33\textwidth]{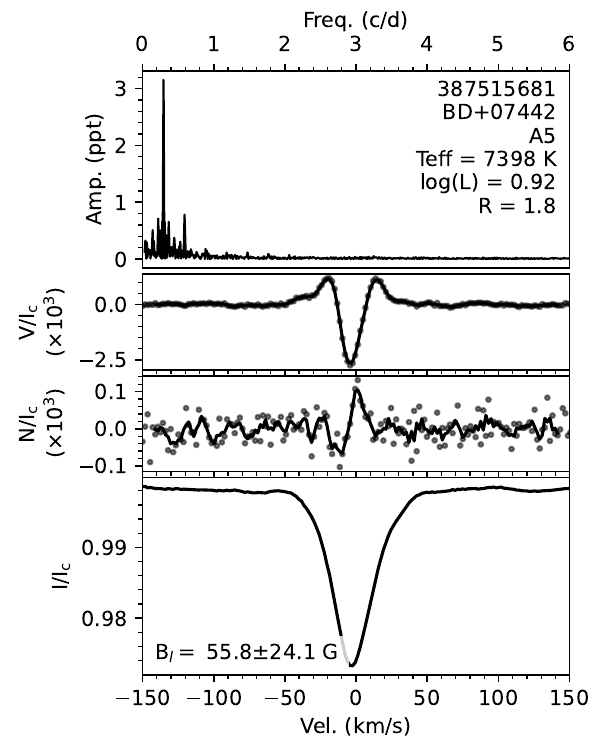}
    \includegraphics[width=0.33\textwidth]{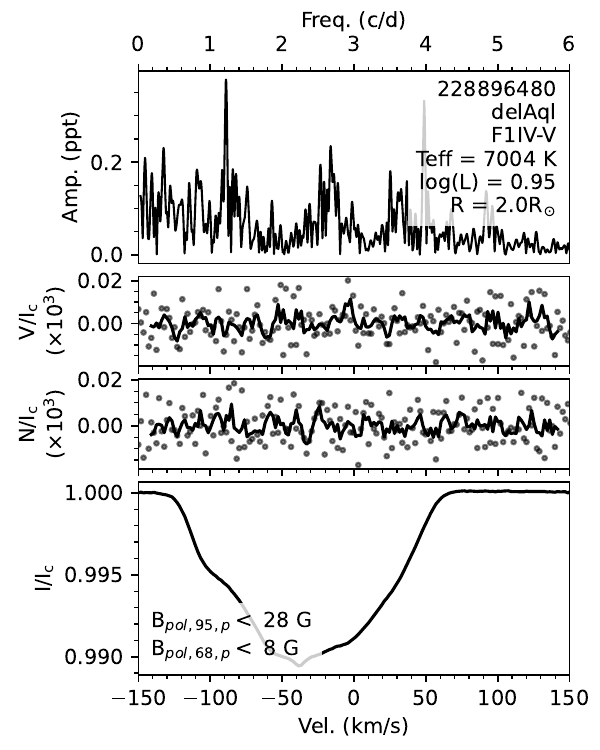}
    \includegraphics[width=0.33\textwidth]{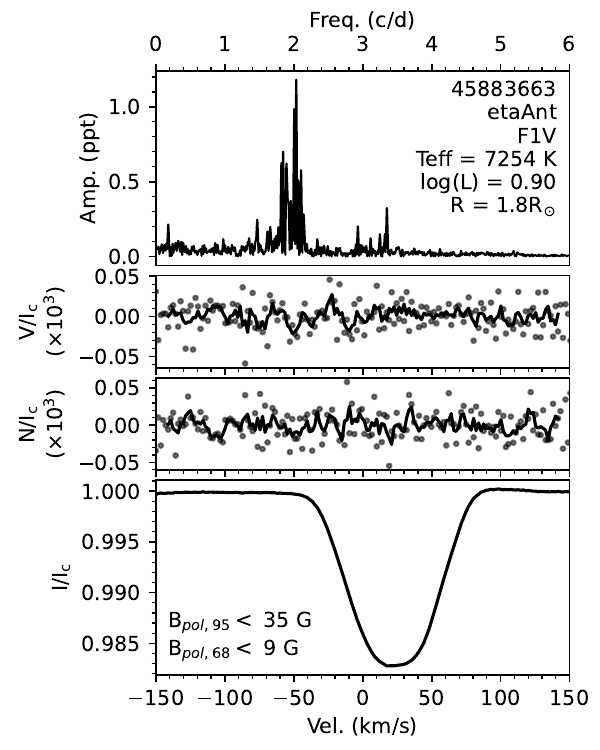}
    \includegraphics[width=0.33\textwidth]{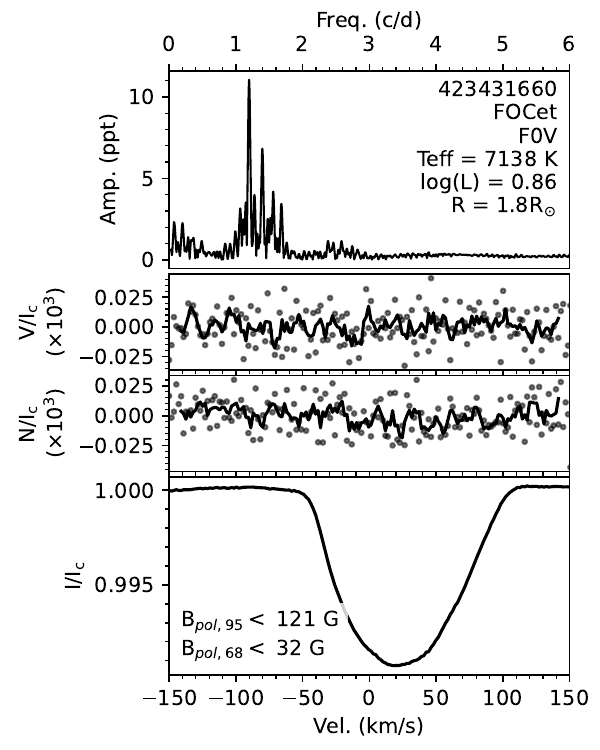}
    \includegraphics[width=0.33\textwidth]{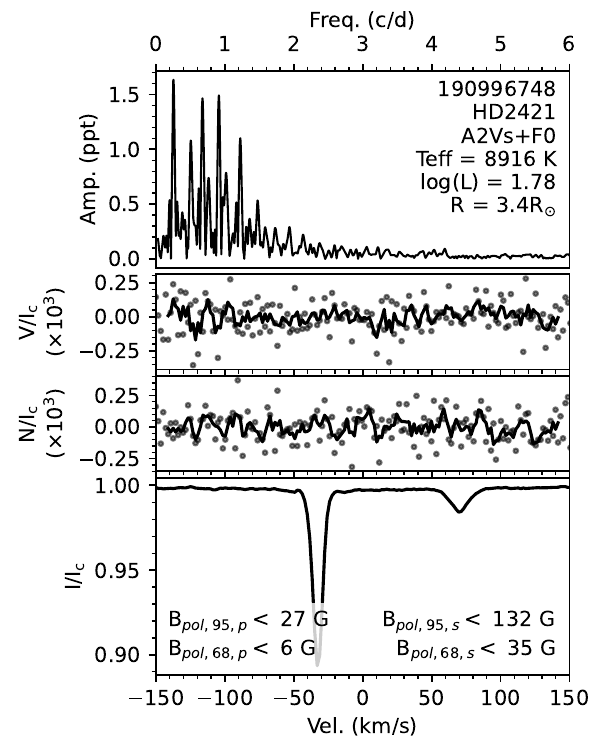}
    \includegraphics[width=0.33\textwidth]{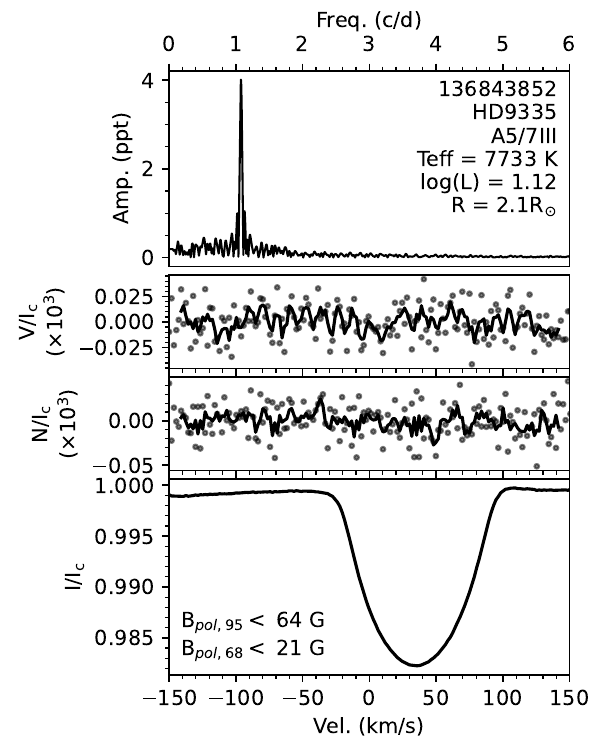}
    \caption{}
    \label{fig:LSDFT_test2}
\end{figure*}

\renewcommand{\thefigure}{C.\arabic{figure} (Cont.)}
\addtocounter{figure}{-1}

\begin{figure*}
    \centering
    \includegraphics[width=0.33\textwidth]{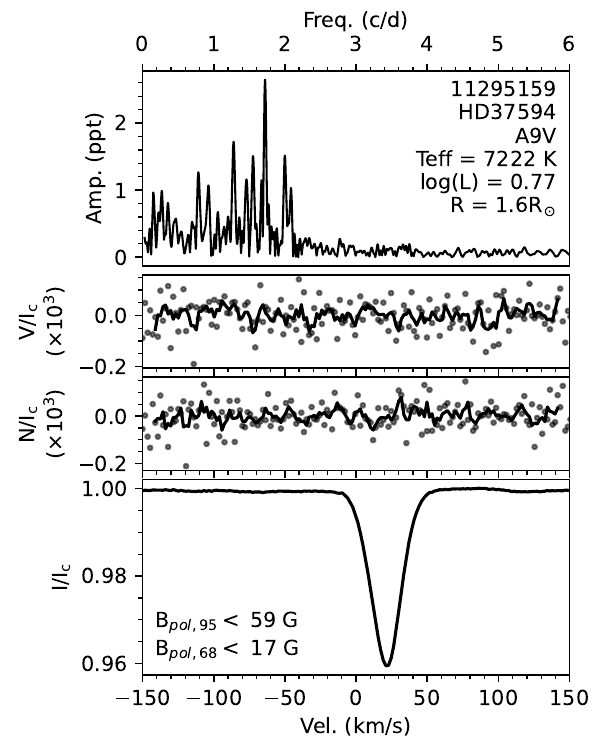}
    \includegraphics[width=0.33\textwidth]{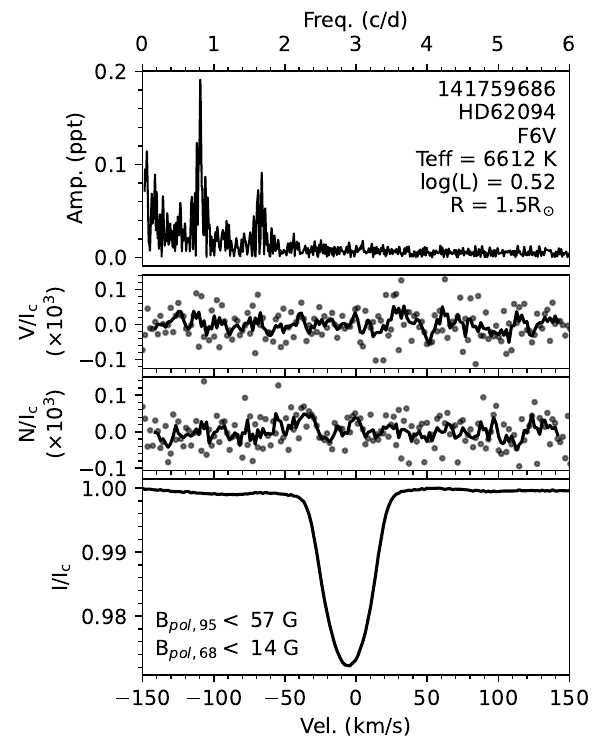}
    \includegraphics[width=0.33\textwidth]{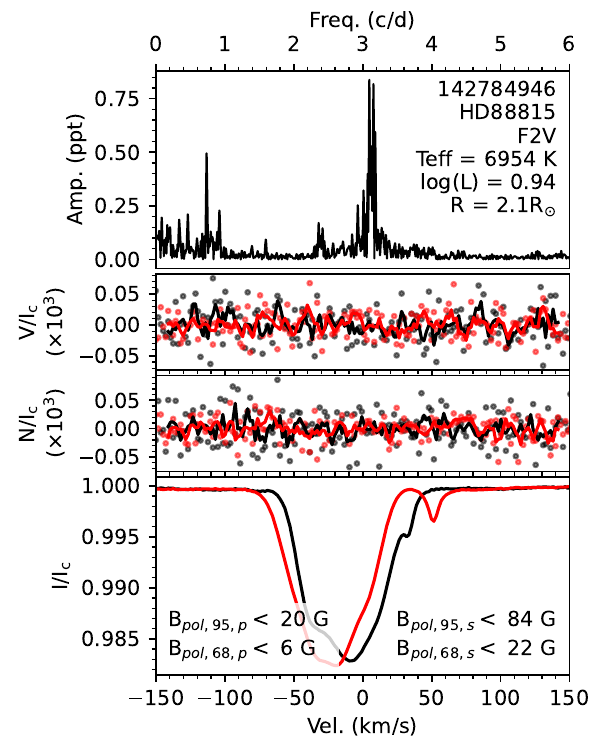}
    \includegraphics[width=0.33\textwidth]{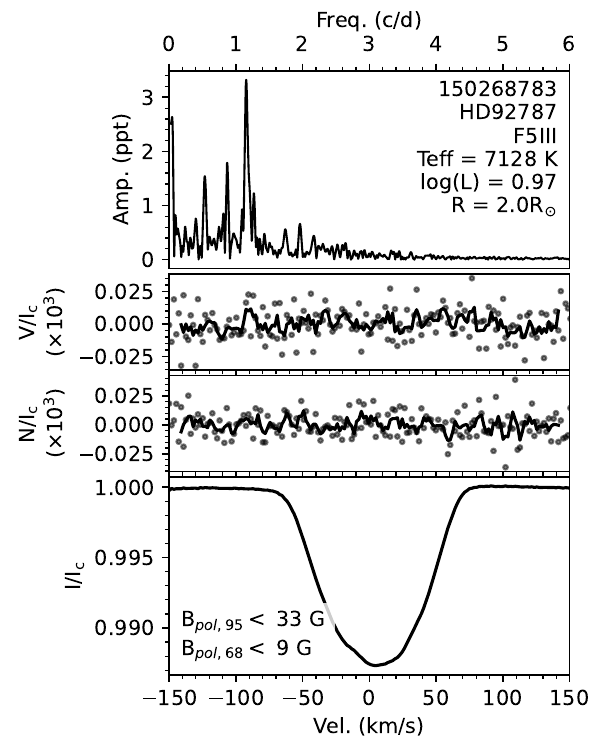}
    \includegraphics[width=0.33\textwidth]{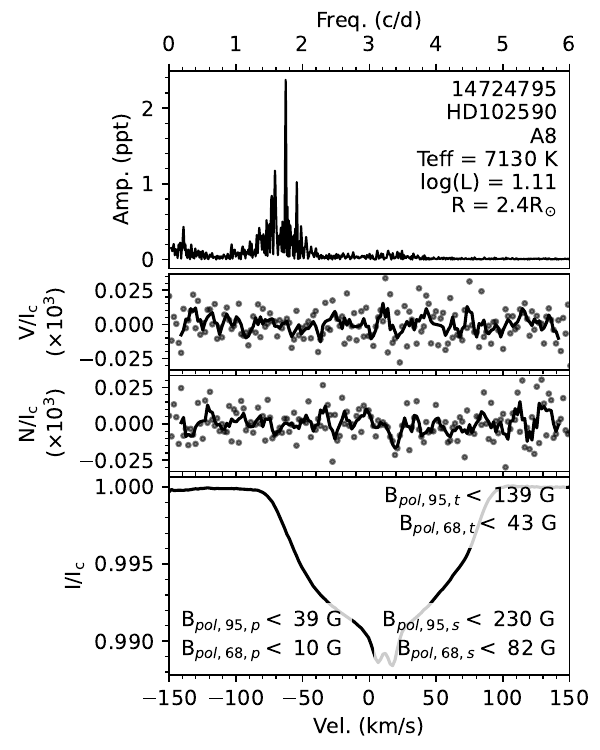}
    \includegraphics[width=0.33\textwidth]{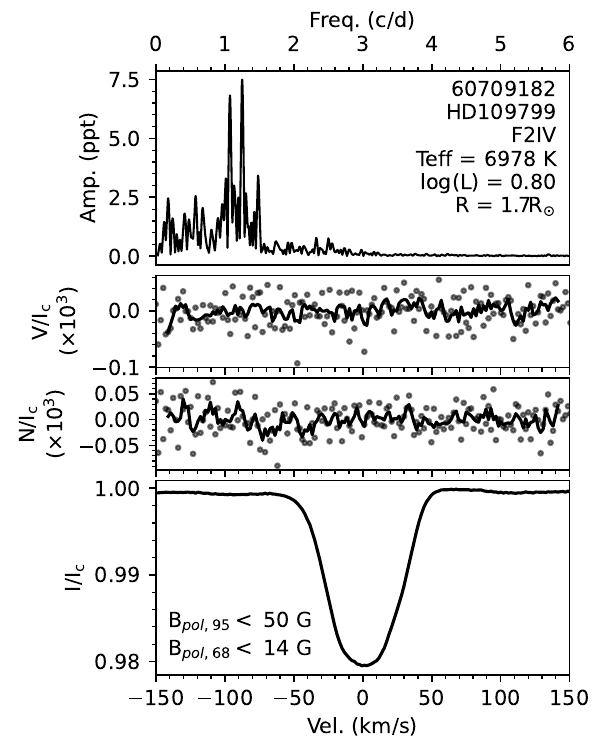}
    \includegraphics[width=0.33\textwidth]{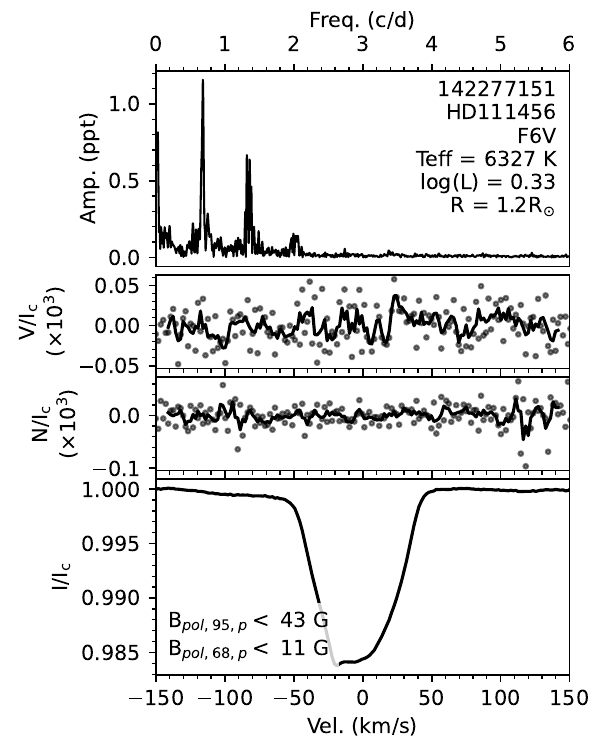}
    \includegraphics[width=0.33\textwidth]{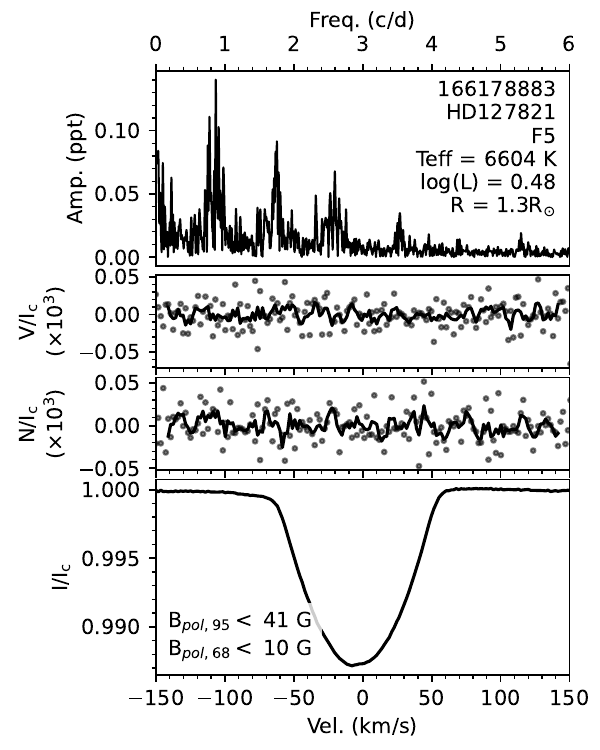}
    \includegraphics[width=0.33\textwidth]{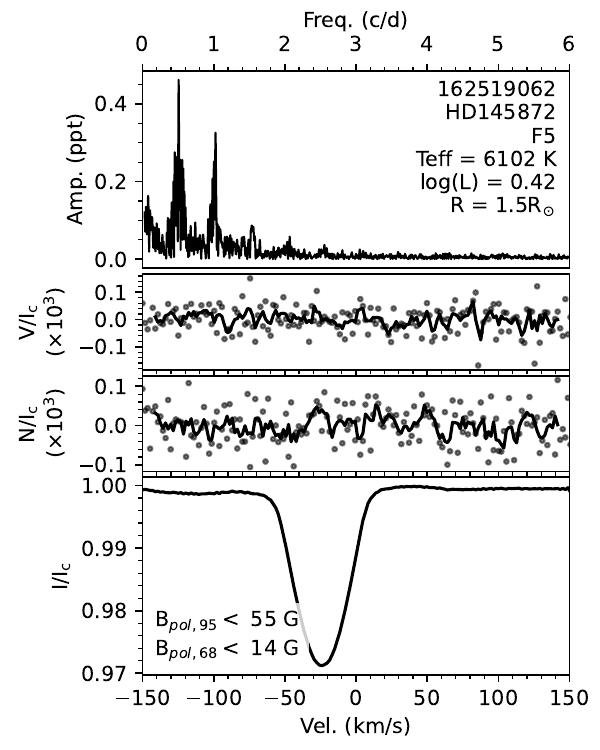}
    \caption{}
    \label{fig:LSDFT_test3}
\end{figure*}

\renewcommand{\thefigure}{C.\arabic{figure} (Cont.)}
\addtocounter{figure}{-1}

\begin{figure*}
    \centering
    \includegraphics[width=0.33\textwidth]{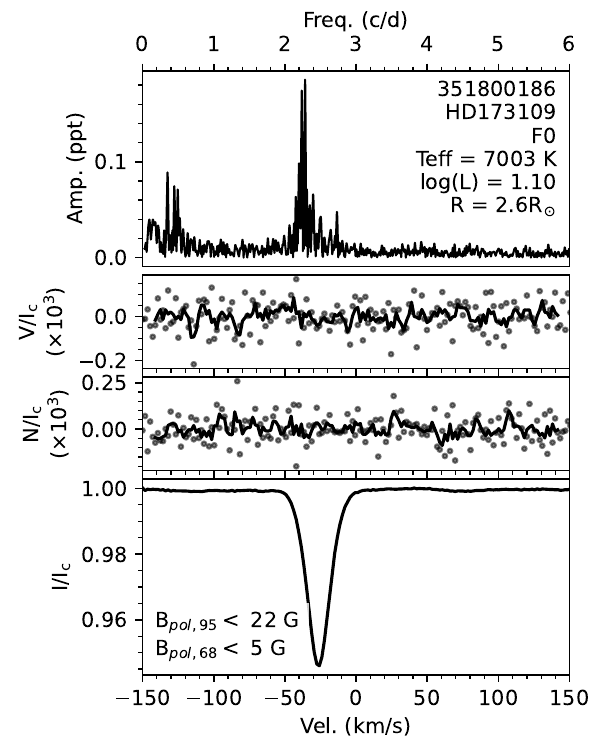}
    \includegraphics[width=0.33\textwidth]{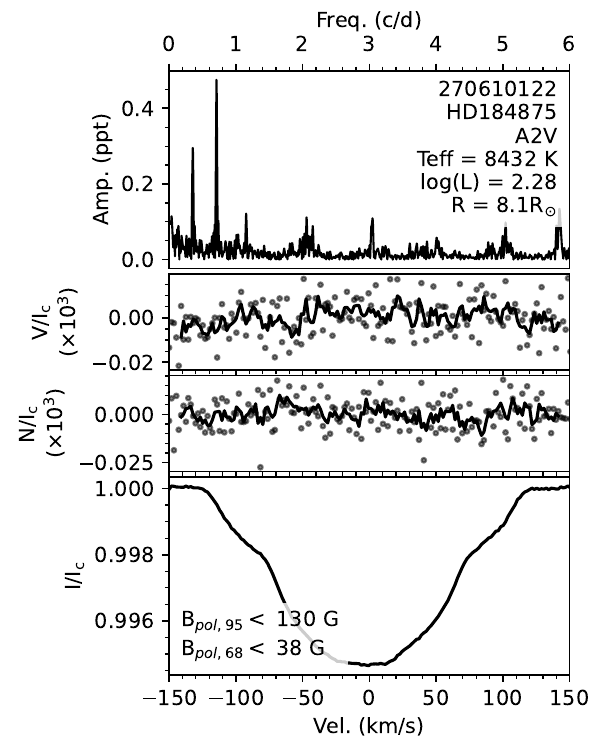}
    \includegraphics[width=0.33\textwidth]{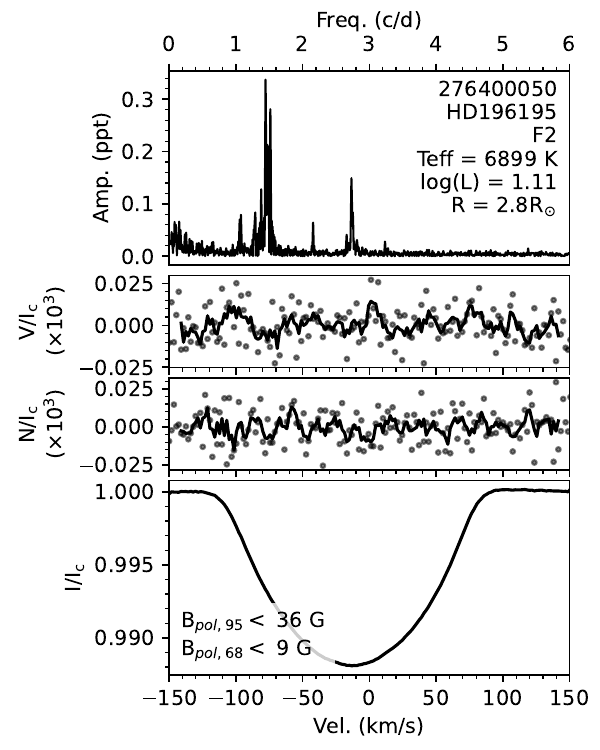}
    \includegraphics[width=0.33\textwidth]{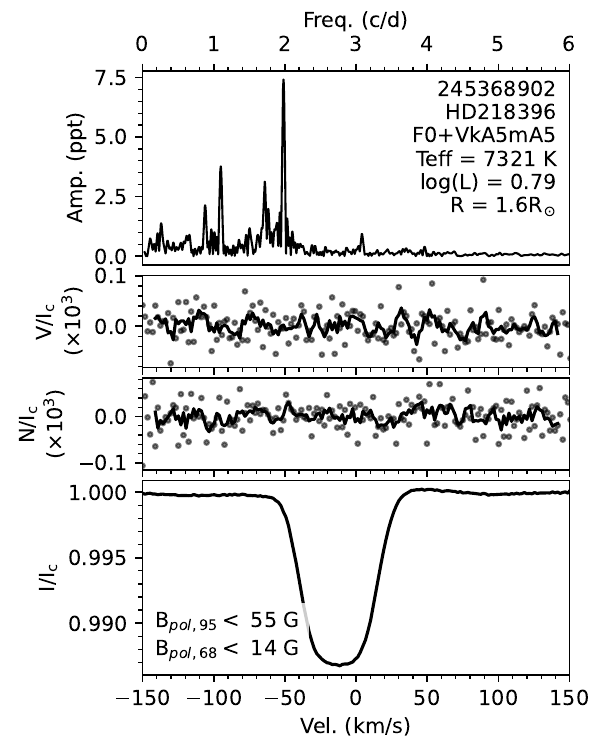}
    \includegraphics[width=0.33\textwidth]{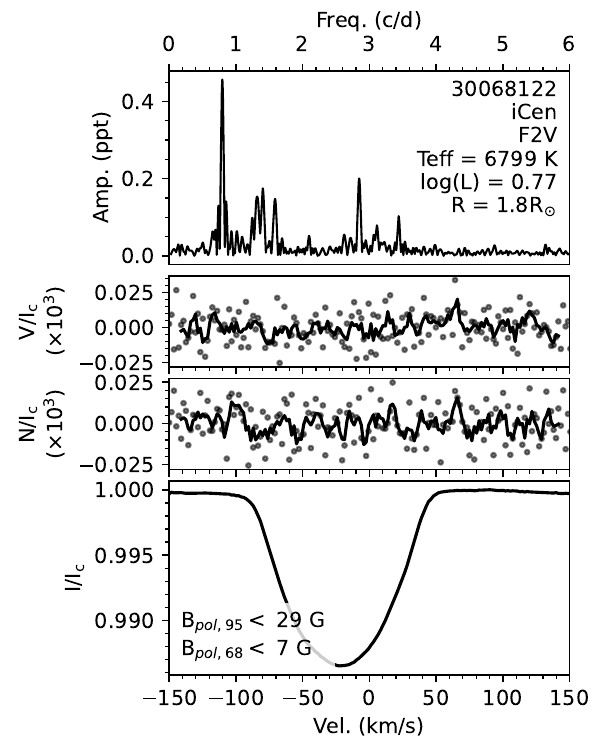}
    \includegraphics[width=0.33\textwidth]{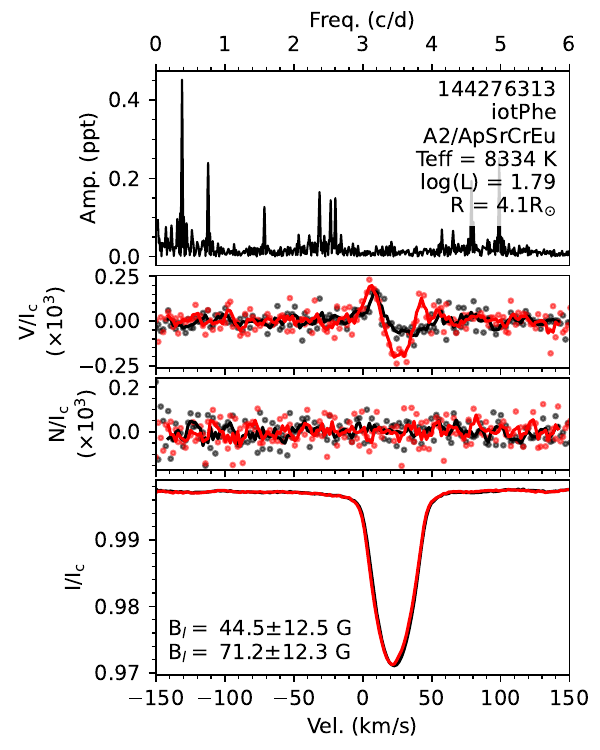}
    \includegraphics[width=0.33\textwidth]{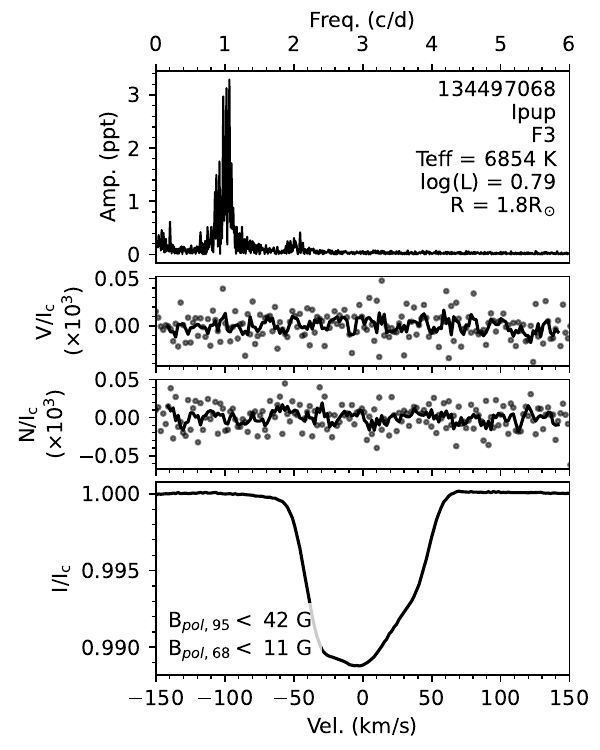}
    \includegraphics[width=0.33\textwidth]{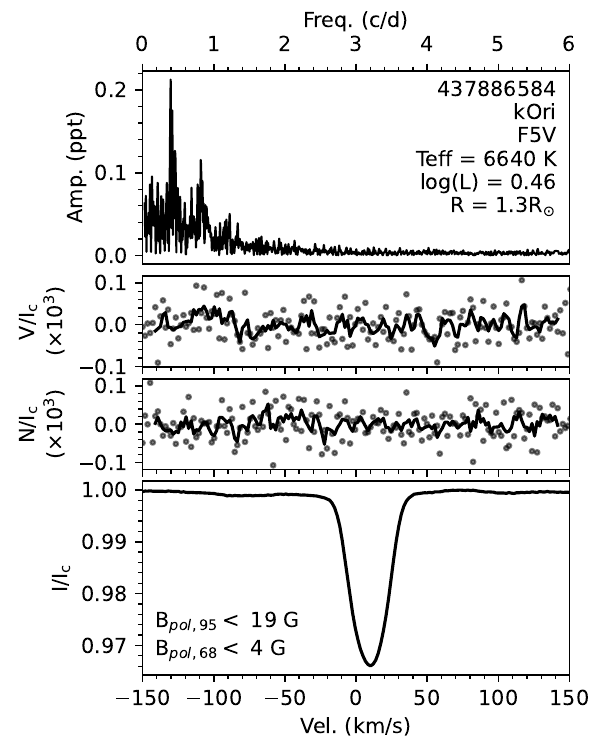}
    \includegraphics[width=0.33\textwidth]{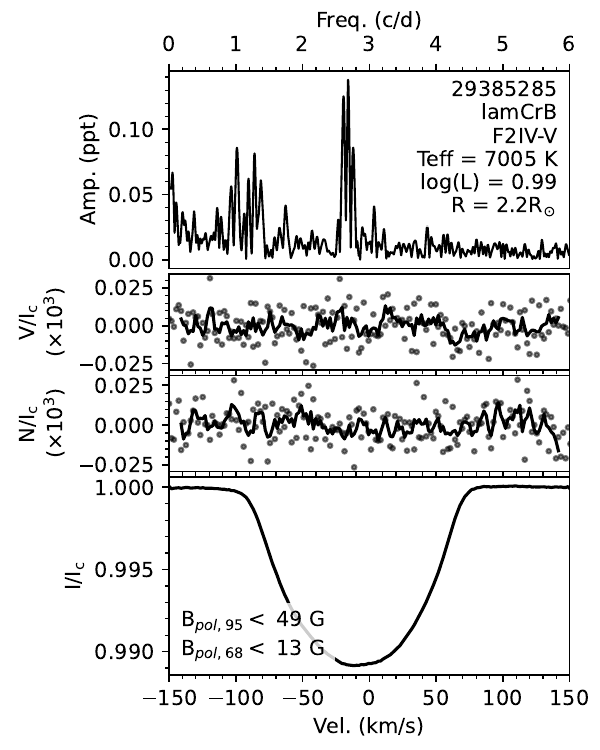}
    \caption{}
    \label{fig:LSDFT_test4}
\end{figure*}

\renewcommand{\thefigure}{C.\arabic{figure} (Cont.)}
\addtocounter{figure}{-1}

\begin{figure*}
    \centering
    \includegraphics[width=0.33\textwidth]{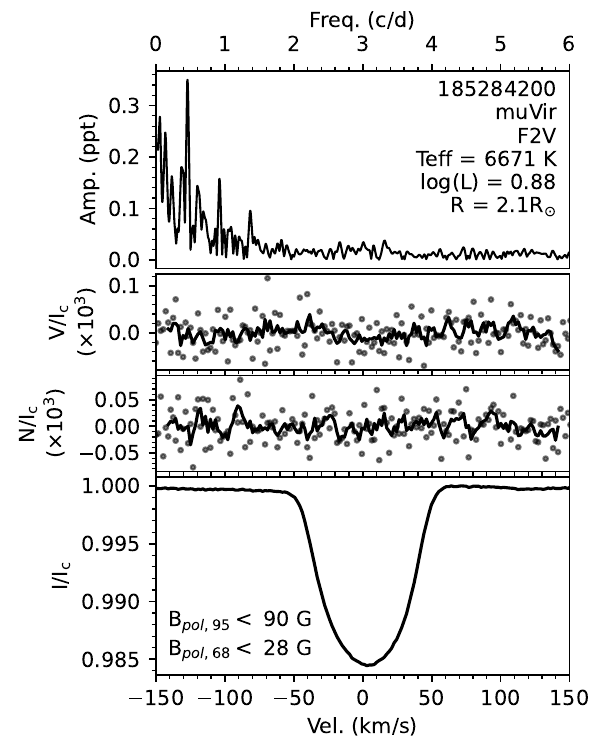}
    \includegraphics[width=0.33\textwidth]{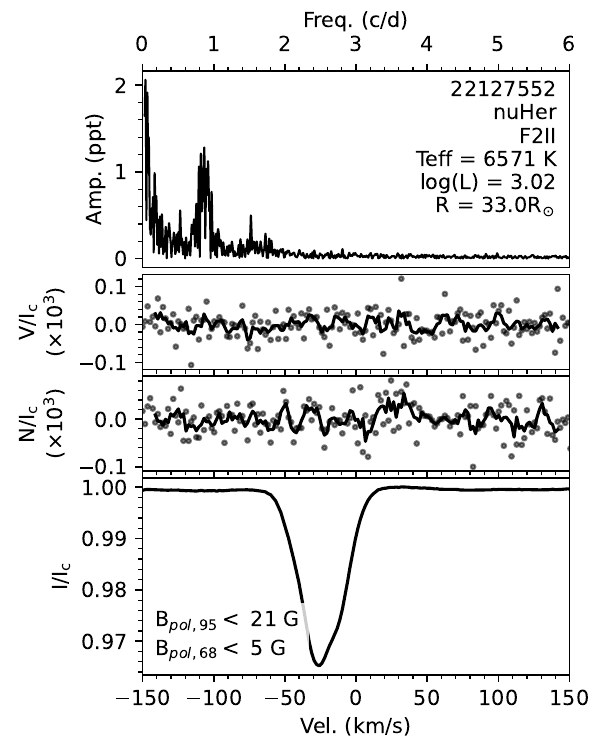}
    \includegraphics[width=0.33\textwidth]{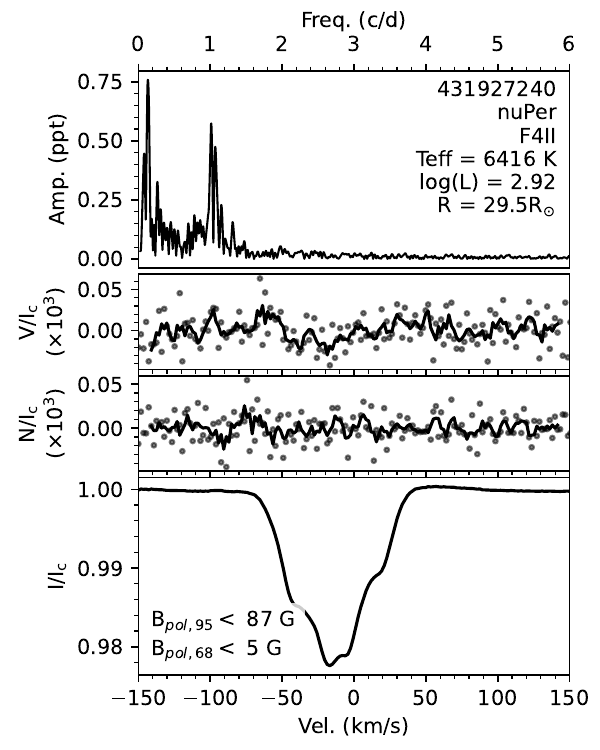}
    \includegraphics[width=0.33\textwidth]{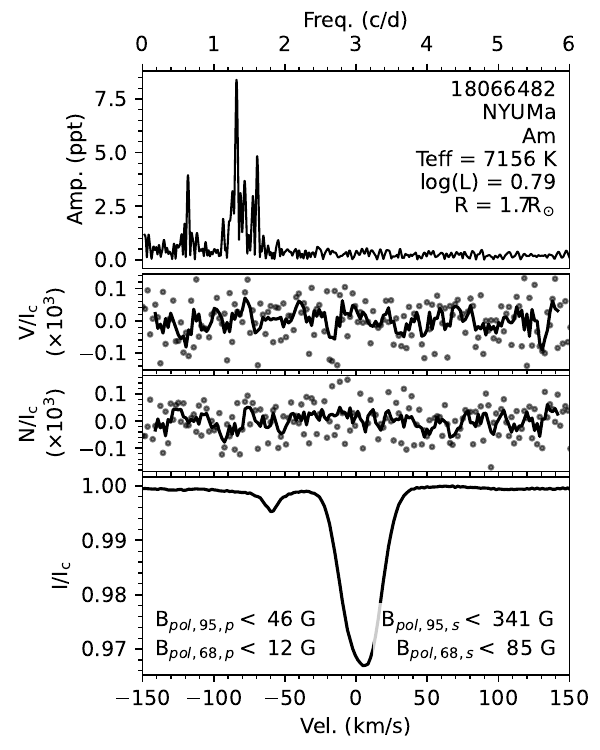}
    \includegraphics[width=0.33\textwidth]{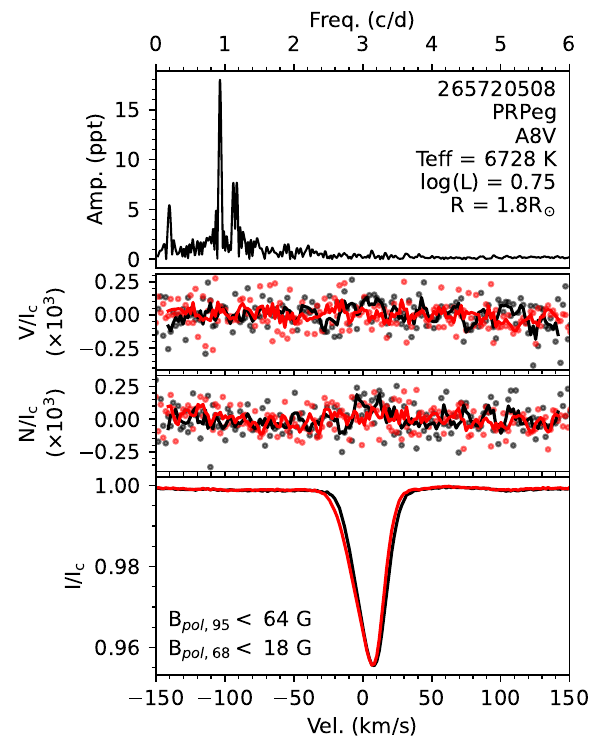}
    \includegraphics[width=0.33\textwidth]{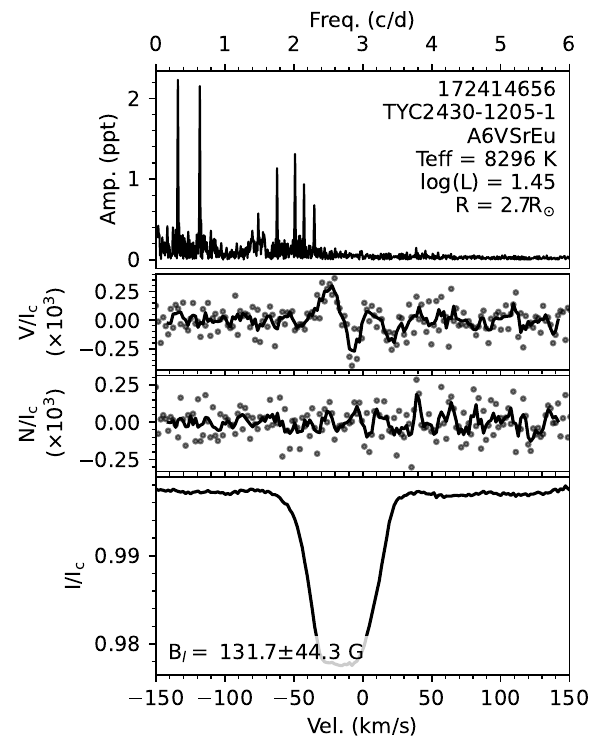}
    \includegraphics[width=0.33\textwidth]{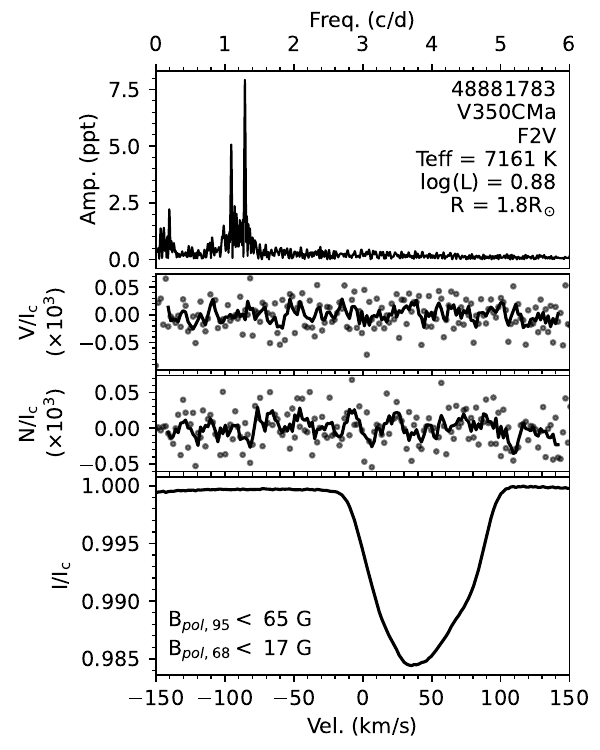}
    \includegraphics[width=0.33\textwidth]{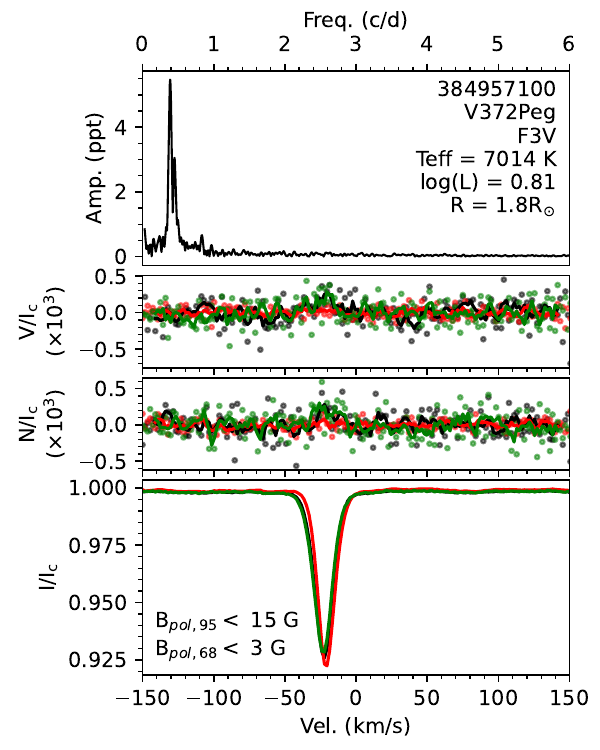}
    \includegraphics[width=0.33\textwidth]{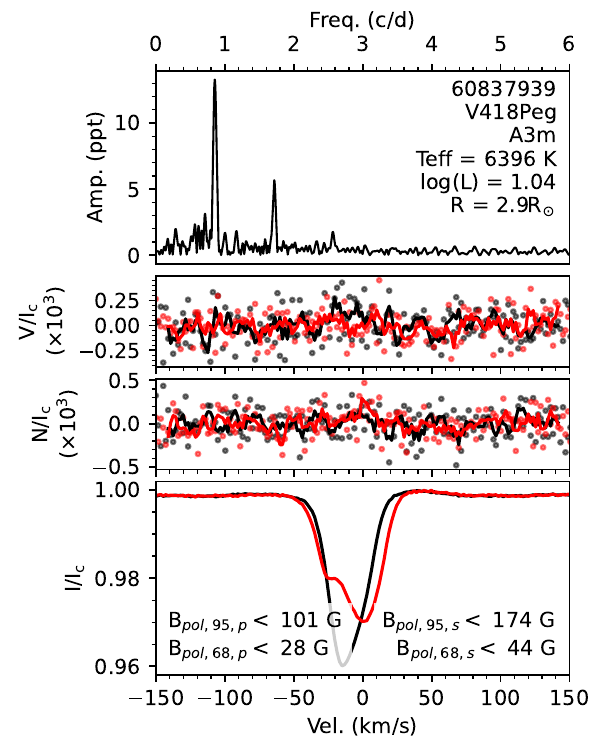}
    \caption{}
    \label{fig:LSDFT_test5}
\end{figure*}

\renewcommand{\thefigure}{C.\arabic{figure} (Cont.)}
\addtocounter{figure}{-1}

\begin{figure*}
    \centering
    \includegraphics[width=0.33\textwidth]{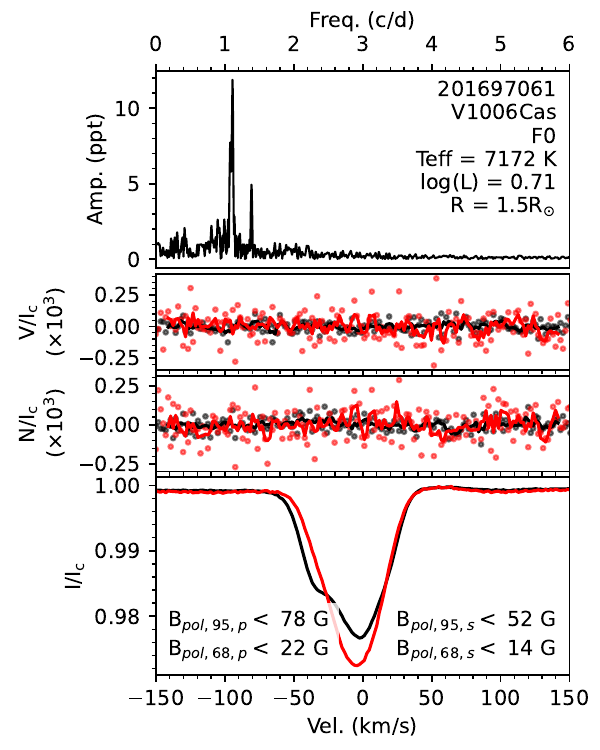}
    \includegraphics[width=0.33\textwidth]{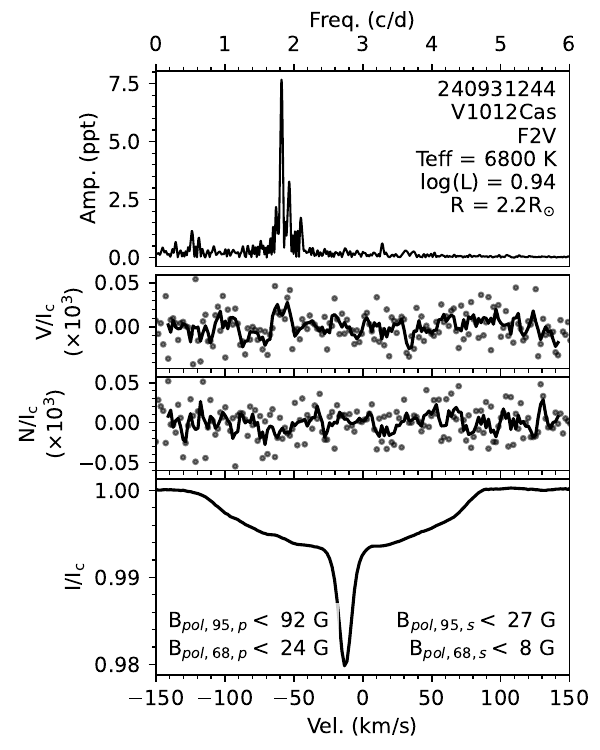}
    \includegraphics[width=0.33\textwidth]{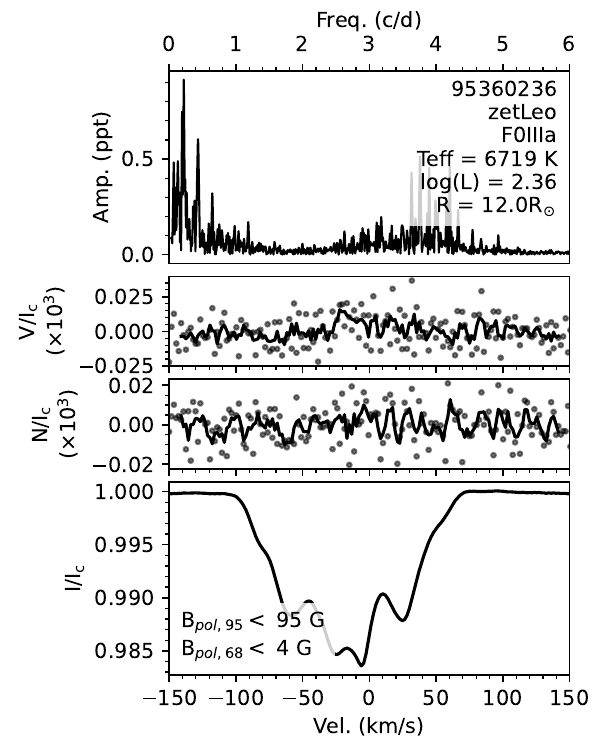}
    \caption{}
    \label{fig:LSDFT_test6}
\end{figure*}

\renewcommand{\thefigure}{D.\arabic{figure}}

\clearpage
\section{BD+07 442 (= TIC 387515681), a non-pulsating magnetic star} \label{sec:mag_bd+07442}

\begin{figure*}[ht]
    \centering
    \includegraphics[width=0.95\textwidth]{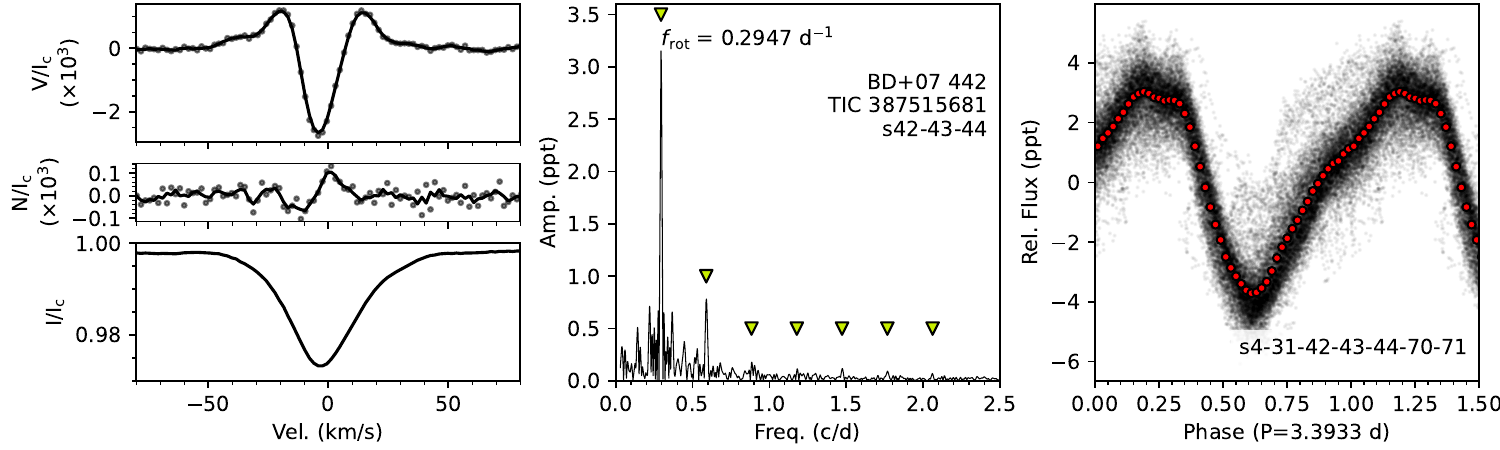}
    \caption{Similar to Fig.~\ref{fig:LSD_iotphe} for the magnetic star BD+07 442. The TESS photometry is phased to the rotation period in the right-most panel.}
    \label{fig:LSD_bd+07442}
\end{figure*}

Figure~\ref{fig:LSD_bd+07442} shows a very strong signal in Stokes $V$ for BD+07~442. This target was one of the earliest observed stars in our sample, with observations being carried out prior to the availability of TESS photometry. Despite the magnetic detection, analysis of TESS data show that this star exhibits only rotational modulation typical of magnetic A-type stars. No pulsational signals are evident. The Stokes N profile is not flat, most likely due to crosstalk from the very strong Stokes V signature -- the reality of the Stokes V signature is not in doubt given its extremely high signal-to-noise ratio, being $\sim$30 times higher amplitude than the variation in the null profile.

To approximate the location of BD+07 442 in the HR diagram of Fig.~\ref{fig:HRD}, we adopt the parameters determined from low-resolution Large Sky Area Multi-Object Fiber Spectroscopic Telescope LAMOST spectra\footnote{\url{http://www.lamost.org/dr11/v2.0/}} in \citet{2017AstBu..72...51S} of T$_{\rm eff}$ = 6870 $\pm$ 56 K, $\log g$ = 3.99 $\pm$ 0.32, R = 2.11 $\pm$ 0.4 R$_{\odot}$, log(L) = 0.96 $\pm$ 0.2.
This is the one object in our sample where the SED fitting process failed to return any results, due to a lack of any parallax information (there is no parallax reported in Gaia DR2 or DR3, nor by Hipparcos). The spectral fitting produced poor fits to the data. There are two likely reasons for this. The first is that this object has been reported as a visual double, where the secondary is approximately one magnitude fainter and located at a distance of $\sim$0.3 arcseconds \citep{2002yCat.1274....0D, 2019AJ....158...48T}. At this small separation, light from both sources was recorded in our spectropolarimetric observations (as well as the TESS photometry). While the Stokes I profile in Fig.~\ref{fig:LSD_bd+07442} is not obviously that of an SB2, the profile is distinctly non-Gaussian (having relatively wide wings), and at large orbital separation the relative RV motion between the two components is expected to be quite small. As an A-type star with a strong magnetic field, the magnetic component is likely chemically peculiar. Indeed, the depression at 5200 $\AA$, which is a hallmark of Ap stars \citep{maitzen76, paunzen05, kochukhov05}, is evident in the LAMOST spectra \citep{2012RAA....12..723Z}. Since our spectral analysis did not take into account the complications owing to chemical peculiarity nor binarity, most regions of the optical spectrum are not well reproduced by synthetic spectra of single chemically-normal stars.  

\section{Stellar parameter determination} \label{apx:spec_SED_fitting}

\renewcommand{\thefigure}{E.\arabic{figure}}

\begin{figure*}
    \centering
    \includegraphics[width=0.4\textwidth]{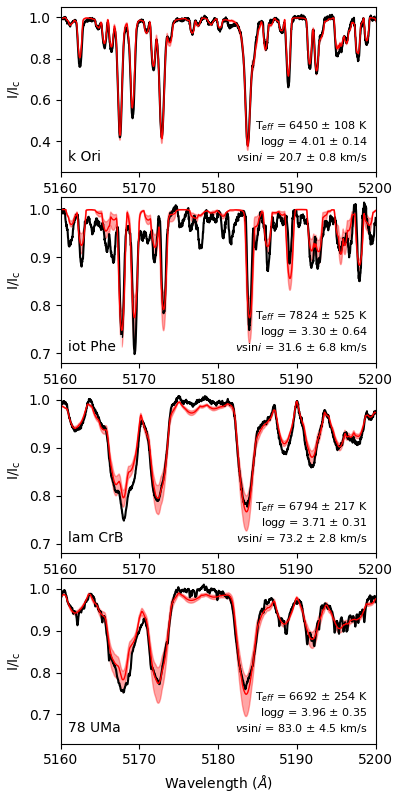}
    \includegraphics[width=0.4\textwidth]{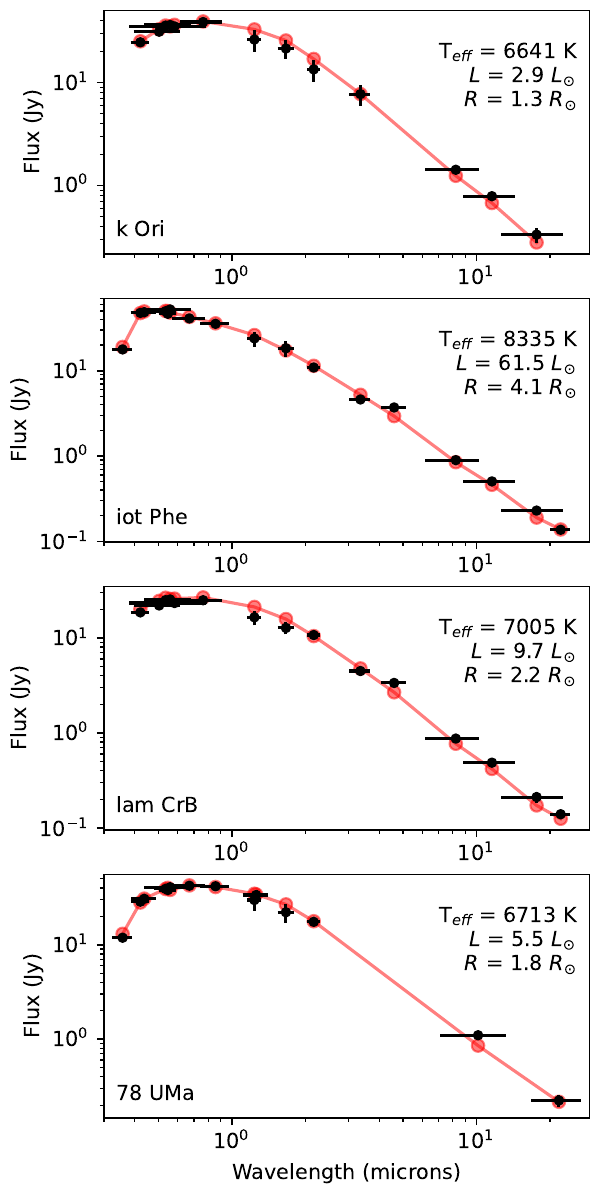}
    \caption{\textit{Left:} Comparison of part of the observed spectrum (black) and the spectral model with the best-fit parameters (red). The shaded red region indicates the uncertainties in $T_{\rm eff}$. \textit{Right:} Comparison of broad-band flux (black) and the best-fit SED points (red). }
    \label{fig:synthspec_fit}
\end{figure*}

Examples of the spectral fitting and SED fitting routines described in Sect.~\ref{sec:stellar_parameters} are shown in Fig.~\ref{fig:synthspec_fit}.
These four objects were chosen as representative examples -- k Ori is a slow rotator, $\iota$ Phe is chemically peculiar and magnetic, $\lambda$ CrB has a high $v \sin i$, and 78 UMa is an SB2 with a rapidly rotating primary and a faint narrow-lined companion. 

\end{appendix}
\end{document}